\pdfoutput=1
\documentclass[sigconf]{acmart}
\renewcommand\footnotetextcopyrightpermission[1]{}

\usepackage{enumitem}
\usepackage{xspace}
\usepackage{graphicx}
\usepackage{subcaption}

\newenvironment{packeditemize}{
\begin{itemize}
  \setlength{\itemsep}{0.3pt}
  \setlength{\parskip}{2pt}
  \setlength{\parsep}{0pt}
}{\end{itemize}}

\newcommand{\sys}{Ada\-Bool\xspace}
\newcommand{\rev}[1]{#1}
\begin{document}
\title{Ada\-Bool: Sparse Boolean Matrix Processing with Adaptive Data Representation}

\author{Xiangrui Yang}
\authornote{Both authors contributed equally to this research.}
\affiliation{
  \institution{The University of Hong Kong}
  \country{Hong Kong SAR, China}
}
\email{yxr620@gmail.com}

\author{Yifan Lei}
\authornotemark[1]
\affiliation{
  \institution{Tencent Inc.}
  \country{China}
}
\email{yifanlei@tencent.com}

\author{Jiawei Yang}
\affiliation{%
  \institution{Huazhong University of Science and Technology}
  \country{China}
}
\email{wadeyoung718@gmail.com}

\author{Yekang Zhan}
\affiliation{
  \institution{Huazhong University of Science and Technology}
  \country{China}
}
\email{zhanyekang@hust.edu.cn}

\author{Yuxin Zheng}
\affiliation{%
  \institution{Tencent Inc.}
  \country{China}
}
\email{yuxinzheng@tencent.com}

\author{Lei Mao}
\affiliation{%
  \institution{Tencent Inc.}
  \country{China}
}
\email{ivanmao@tencent.com}

\author{Zhaoren Wu}
\affiliation{%
  \institution{Tencent Inc.}
  \country{China}
}
\email{jasonzrwu@tencent.com}

\author{Yibo Huang}
\affiliation{%
  \institution{University of Michigan}
  \country{USA}
}
\email{ybhuang.cs@gmail.com}

\author{Yiming Qiu}
\affiliation{%
  \institution{The University of Hong Kong}
  \country{Hong Kong SAR, China}
}
\email{yimingq@hku.hk}

\begin{abstract}
Many data processing tasks revolve around determining whether logical relationships exist between two sets of entities, a pattern that can be naturally modeled as \textit{Sparse Boolean Matrix Multiplication (SpBMM)}. Under Boolean semantics, matrix representations no longer need to store numerical values and instead focus on encoding the positions of non-zero elements,
making it highly amenable to data compression for substantially lower memory access overhead.
The key challenge, however, is that the most effective compression depends heavily on sparsity distribution: different methods perform best at different sparsity levels, while real-world matrices often contain sharply varying sparsity across regions. Consequently, forcing the entire matrix into any fixed, monolithic representation leaves considerable optimization opportunities untapped.

We present \sys, a SpBMM framework designed to reconcile this tension through an adaptive data representation.
\sys partitions matrices into fine-grained element blocks and represents each block with the encoding best matched to its local sparsity.
On top of this adaptive encoding, \sys introduces a heterogeneous GPU execution model that groups blocks by encoding type and maps them to matching processing paths, enabling efficient execution and scheduling across GPU resources. We implement \sys end to end with low-level system optimizations, evaluate it on both production and public workloads, and show that it outperforms representative baselines by over an order of magnitude.
\end{abstract}

\maketitle
\pagestyle{plain}

\vspace{-1mm}
\section{Introduction}

Many data processing tasks in modern systems ultimately ask a simple question: what relationships exist between two sets of entities? This question arises across a range of settings, including filtering and matching in retrieval and recommendation systems~\cite{borisyuk2024linr,liu2018griffin,terekhov2020context, shen2024learning}, as well as reachability analysis in graph databases \cite{ai2024Boolean}. A natural way to express such workloads is to encode entities and their relationships as matrices and vectors, turning relationship evaluation into a matrix processing problem. In practice, these matrices are often both sparse and Boolean: most possible relationships do not exist, and the computation only needs to determine existence rather than compute a numerical value \cite{wang2023tc, mouhah2023large}. This makes \textit{Sparse Boolean Matrix Multiplication (SpBMM)} a natural abstraction for a wide range of data processing tasks, as it reduces computation in general \textit{Sparse Matrix Multiplication (SpMM)} \cite{buono2016optimizing, wang2021gnnadvisor} to simple Boolean operations. As modern data systems scale to millions of entities, efficient SpBMM becomes increasingly critical for sustaining high efficiency in production environments \cite{borisyuk2024linr, ai2024Boolean, douze2025faiss, johnson2019billion}.

\rev{
Today, GPUs are widely employed to accelerate data processing tasks, as they offer abundant parallelism for matrix computations~\cite{nagasaka2017high, 10.1145/3571157, wang2021comprehensive, shanbhag2020study}.
}
To exploit matrix sparsity, GPU computation is formulated not on dense arrays, but on encoded sparse structures that bypass zero-valued positions. For this reason, \textit{data representation} is fundamental to SpMM performance: whatever execution or memory optimization a system applies, it must ultimately build on some assumption about how sparse structures are encoded and accessed. This dependence becomes even stronger in SpBMM, where computation becomes cheaper as numerical arithmetic is replaced by lightweight Boolean processing, yet the overhead of accessing sparse structures remains.
As such, SpBMM is inherently a \textit{memory-bound} workload \cite{mei2016dissecting}. A data representation that can be accessed efficiently across the GPU memory hierarchy therefore becomes crucial for the end-to-end performance of SpBMM systems.

Prior sparse matrix processing systems largely share a common design assumption: they commit to a fixed, monolithic data representation to encode entire matrices~\cite{kobus2018cubool, gale2020sparse, naumov2010cusparse, chen2022bit, chen2023gpu}. General SpMM frameworks, for example, are commonly built on formats such as Compressed Sparse Row (CSR), which store non-zero values and their indices in dedicated arrays \cite{greathouse2014efficient}. Existing works then layer increasingly sophisticated scheduling strategies and memory access optimizations on top of this representation in an effort to improve performance \cite{naumov2010cusparse, gale2020sparse, wang2023tc, parger2020speck}. However, once the representation itself is fixed, these optimizations can only explore the design space constrained by that choice. This restriction is especially important for SpBMM, as the following observations illustrate.

\vspace{-1mm}
\setlist[itemize]{leftmargin=*}
\begin{packeditemize}

\item \textbf{Opportunities for compression.}
Since elements in Boolean matrices denote logical existence rather than numerical quantities, the main burden of data representation shifts from storing element values to encoding the positions of non-zero entries \cite{kobus2018cubool}. This opens the door to \textit{lossless data compression} \cite{wang2017experimental, chambi2016better}, using either (1) an \textit{offset array} that stores relative offsets of indices rather than their absolute positions, or (2) a \textit{bitmap} format that encodes a contiguous range of positions with one bit per position.

\item \textbf{Dependence on sparsity level.}
No compressed data structure is universally optimal; the better choice depends on sparsity.
When a matrix is highly sparse, most positions are zero-valued. In this case, offset-style compression is more effective because it can skip zero-valued positions, whereas a bitmap still allocates one bit to each position in the range. When sparsity is lower and non-zero entries become denser, per-entry compression starts to dominate instead. In this case, bitmaps become more efficient than offset arrays because they encode each position with a single bit.

\item \textbf{Skewness in sparsity distribution.}
Boolean matrices in real world settings usually come with \textit{highly-skewed sparsity distributions} \cite{yang2016write, cristelli2012there}, which means different regions of the same matrix may present distinct sparsity levels. In other words, there is no one-size-fits-all data structure even if we know the characteristics of the matrix a priori.
Strong performance lies in adapting the choice of compression method based on the sparsity of fine-grained matrix regions, rather than that of entire matrices.

\end{packeditemize}
\vspace{-1mm}

These properties suggest that sparse Boolean matrices should not be forced into a monolithic representation with a single data structure. Instead, it calls for an \textit{adaptive} representation that leverages the strengths of different data structures at the granularity of matrix regions. Realizing this idea hinges on answering two questions: \textit{which} data structures should be used, and \textit{when} to use them. To answer the first, we need to devise data structures that are specialized to different sparsity levels, so that together they can cover fine-grained local sparsity variation. To answer the second, we must quantify local sparsity in a way that allows each region to be assigned the most suitable structure. To this end, we introduce the \textit{\sys representation}, an adaptive Boolean data representation for SpBMM. \sys partitions each matrix row into small, fixed-size \textit{element blocks}, evaluates the local sparsity level of each block by counting its non-zero positions, and adaptively encodes each block using either a family of offset-based arrays or a dense bitmap structure, whichever best matches that block’s sparsity level.

To fully exploit the optimization space enabled by the adaptive data representation, the GPU execution must also move beyond a monolithic pass over a fixed format. We therefore organize the \textit{\sys execution model} around three representation-driven principles. First, \sys applies \textit{role-aware encoding}:
matrix encodings are configured according to their role in the computation so that the conversion process of the adaptive data representation
stays off the runtime critical path. Second, \sys adopts \textit{type-based execution}: regions encoded with different data structures are grouped into separate pools and processed by different GPU kernels, keeping each kernel’s workload homogeneous and enabling more efficient, coalesced memory accesses.
Third, building on this type-aware organization, \sys enables \textit{block-level scheduling}: element blocks within the same pool share the same type and size, allowing lightweight, fine-grained scheduling strategies that balance workload across GPU execution units without heavy runtime mechanisms.
Together, these designs turn the heterogeneous \sys representation into an efficient and balanced GPU execution substrate.

We implement \sys data representation and execution model end to end as a practical SpBMM processing system, which further incorporates lower-level system optimizations to matrix memory layout and metadata indexing, reducing irregular accesses and improving data movement efficiency during execution. We evaluate the \sys system through extensive microbenchmarking, validation on real-world datasets from both our production environment and public sources, comparisons against state-of-the-art systems, and detailed ablation analyses. Across these experiments, the benefits of adaptive data representation, the GPU execution model, and the system optimizations compound effectively in practice, enabling \sys to outperform prior designs by more than an order of magnitude.
Our key contributions are outlined as follows:

\begin{packeditemize}

\item We characterize real-world SpBMM tasks, including \textit{AdsFilter}, an open-sourced dataset from our production recommendation pipeline, and identify key challenges and opportunities.

\item We design ~\cite{adabool-repo} the \textit{\sys} framework, centered on an adaptive data representation, supported by a heterogeneous execution model with corresponding system optimizations.

\item We implement and evaluate \textit{\sys}, showing substantial performance gains over representative state-of-the-art systems.

\end{packeditemize}

\section{Background}
In this section, we analyze the Boolean matrix multiplication workload and discuss the limitations in existing works.

\subsection{Boolean Matrix Multiplication}
\label{sec:bool_mat}

\begin{figure}[t!]
  \centering
\includegraphics[width=0.79\linewidth]{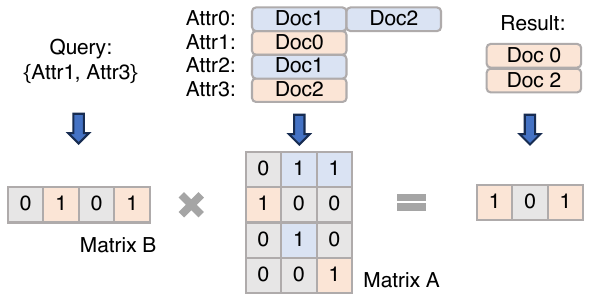}
  \vspace{-3mm}
  \caption{Example of SpBMM: attribute-based filtering.}
  \vspace{-4mm}
  \label{fig:attribute_filtering}

\end{figure}

To understand the application of Boolean matrices in real-world scenarios,
we examine its usage in our attribute-based filtering servers. In this system, an incoming query, which contains a set of attributes, is matched against the attributes associated with each document  \cite{borisyuk2024linr, liu2018griffin}. A document is included in the result if any of its associated attributes appear in the query. This matching process can be expressed using concise Boolean matrix semantics.
As shown in Figure \ref{fig:attribute_filtering},
the document dataset forms a $4 \times 3$ Boolean matrix $A$ where columns are distinct documents and rows are their associated attributes;
The queries form another Boolean matrix $B$, with each row being a query against different attributes. The result of the Boolean matrix multiplication $BA$ \cite{borisyuk2024linr}  gives another matrix $C$, with each row showing documents that contain the queried attributes.

Formally, Boolean matrix multiplication operates over the Boolean semiring $(\{0,1\}, \vee, \wedge)$, replacing standard arithmetic addition and multiplication with logical OR ($\vee$) and logical AND ($\wedge$), respectively \cite{fischer1971Boolean, drakopoulos2022genetic, kobus2018cubool}. For the two Boolean matrices $A$ and $B$, their product $C = BA$ is defined element-wise as:

\vspace{-2.5mm}
$$
(C)_{ij}=\bigvee_k(B_{ik}\wedge A_{kj})
$$
\vspace{-2.5mm}

 \noindent In many practical deployments, one operand $A$ (e.g., membership tables) is static and reused across many queries, whereas $B$ (e.g., membership queries) and the resulting $C$ are short-lived inputs and outputs. As a result, the choice of data representation must be weighed against its conversion overhead differently across operands.

\begin{figure}[t!]
  \centering
  \includegraphics[width=\linewidth]{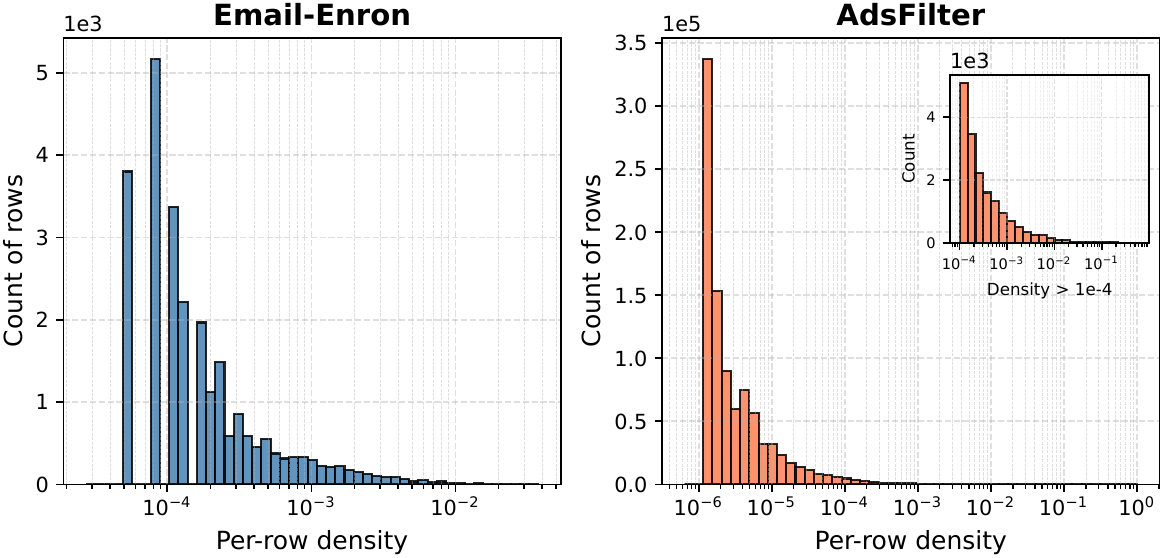}
  \vspace{-7mm}
  \caption{Per-row density (percentage of non-zero elements per row) for representative public (Email-Enron) and production (AdsFilter) SpBMM datasets.}
  \Description{Email-Enron Dataset (left). AdsFilter dataset (right) 6,899,568 * 1,554,642.}
  \vspace{-3mm}
  \label{fig:row_density}
\end{figure}

In real-world scenarios, Boolean matrices often encode membership or incidence over very large universes (e.g., millions of items), where each row touches only a tiny fraction of possible connections
\cite{kuramochi2005finding, kimmerer2024graphblas}.
Consequently, Boolean matrices
are usually extremely sparse \cite{ai2024Boolean, borisyuk2024linr, yang2022graphblast, zhang2018surf}.
In fact, the density of Boolean matrices, defined as the ratio of the number of non-zero elements to the total number of elements, is often around $10^{-4}$.
This is a level of sparsity that is more significant than that of numerical matrices ~\cite{gale2020sparse}.
Furthermore, the sparsity of Boolean matrices exhibits significant skewness.
As shown in Figure~\ref{fig:row_density}, both public datasets like email network (Email-Enron~\cite{leskovec2009community}, \rev{size $36{,}692 \times 36{,}692$}) and the attribute-based filtering dataset (AdsFilter, \rev{size $6{,}899{,}568 \times 1{,}554{,}642$}), from our production recommendation system, exhibit highly non-uniform row densities, ranging from $10^{-6}$ to $10^{-2}$.
This happens because in many real-world domains, a small number of entities, such as hub nodes in communication networks or popular attributes in filtering systems, are connected to disproportionately many others, resulting in their rows containing significantly more non-zero elements than those in the long tail.
This phenomenon leads to highly skewed row-wise distributions
\cite{yang2016write, cristelli2012there}.

\subsection{Sparse Matrix Multiplication in GPUs}
\label{sec:exist}
Graphics Processing Units (GPUs), which commonly adopt Single Instruction, Multiple Threads (SIMT)-style execution model, have become a de facto standard for accelerating large-scale matrix operations \cite{10.1145/3571157}. The prevalent programming model for GPUs is CUDA, which organizes computational tasks into a hierarchy of thread blocks and warps (groups of 32 synchronously executing threads) \cite{abdelfattah2016performance}.
In GPU-based \rev{sparse} matrix multiplication, a substantial amount of execution time and memory bandwidth is used to load matrix data from global memory into thread blocks. In fact, the storage and indexing of matrix data frequently take precedence over computation operations in terms of overall performance.
To efficiently store and index sparse matrices on GPUs, standardized data representations such as coordinate format (COO) and compressed sparse row format (CSR) have been proposed and widely adopted \cite{karakasis2012extended, borvstnik2014sparse}. \rev{COO stores one coordinate tuple (row index, column index, value) for each non-zero entry.}
\rev{CSR, on the other hand, eliminates redundant row indices by compressing them into an array of row pointers, recording them alongside column indices and non-zero values.}
Its compact storage format allows for contiguous memory layout
and row-based task partitioning strategies, which have made it a popular choice for general SpMM tasks.

\begin{figure}[t!]
  \centering
  \includegraphics[width=0.95\linewidth]{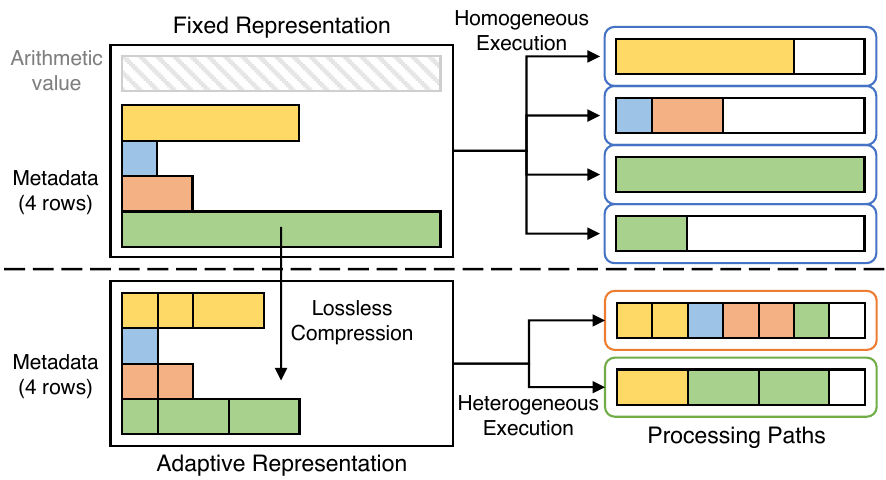}
  \vspace{-3mm}
  \caption{Comparison between existing and envisioned data representations and execution models.}
  \vspace{-5mm}
  \label{fig:representation_comparison}
\end{figure}

Many prior studies have attempted to speed up sparse matrix processing by improving underlying data representations or optimizing the execution strategies built on top of them~\cite{gale2020sparse, naumov2010cusparse}.
As shown by the top half of Figure~\ref{fig:representation_comparison},
\rev{the colored metadata correspond to the original matrix rows and store the column indices of their non-zero entries, while the gray bar above them denotes the separate arithmetic-value array.}
Popular frameworks such as \rev{Sputnik~\cite{gale2020sparse}} and \rev{cuSPARSE~\cite{naumov2010cusparse}} are built on standardized formats like CSR, and adopt row-based partitioning with dynamic scheduling to improve parallelism.
However, since they are designed for general SpMM, they retain arithmetic element values and perform numerical computations, incurring substantial wasted memory access and redundant computation when applied to SpBMM.
Recognizing the special properties of Boolean matrices, recent work has introduced targeted optimizations. For example, \rev{cuBool~\cite{orachev2021spbla}} adopts a reduced CSR format that eliminates the arithmetic value array of CSR.
While this simplification improves space efficiency,
existing solutions still preserve the CSR metadata  (i.e., row and column indices) format and continue to leverage a monolithic data representation with a homogeneous execution model~\cite{nagasaka2017high, terekhov2020context, green2012gpu}, preventing them from exploring metadata compression opportunities and exploiting fine-grained sparsity variation within the same Boolean matrix.

\subsection{Challenges and Opportunities}

\noindent\textbf{Metadata-centric compression.}
In Boolean matrices,  each element only denotes logical existence (i.e., true or false).
False entries can therefore be omitted entirely, while true entries do not require storing an additional arithmetic value.
As a result, SpBMM only needs to store and index metadata, whereas in general SpMM arithmetic element values remain a dominant source of representation cost.
This shift opens the door to lossless metadata compression schemes that can greatly improve memory access efficiency. One option is an \textit{offset-based} encoding, which stores a starting index and then records subsequent non-zero entries as relative offsets instead of full indices. Another is a \textit{bitmap-based} encoding, which represents a contiguous region of the matrix using one bit per position to indicate whether that position is non-zero \cite{chambi2016better}.

\noindent\textbf{Sparsity-aware representation.}
As real-world Boolean matrices often exhibit highly skewed, spatially varying sparsity, compression opportunities differ substantially across matrix regions. Accordingly, while both offset-based and bitmap-based compression methods outperform classical CSR-style data structures, neither is universally optimal. When a region contains many non-zero elements, bitmaps are highly effective because they encode each position with a single bit. However, in extremely sparse regions, bitmaps become less effective because most of their space is spent representing zeros. Offset-based arrays behave in the opposite way: they are very compact when sparsity is high because they naturally skip zero-valued positions, but offer a much lower compression rate than bitmaps when non-zero elements are dense. Together, these observations imply that any single, fixed data structure will be fundamentally suboptimal for some regions of some matrices, and that an effective data representation must combine different data structures to adapt to different sparsity levels.

\noindent\textbf{Heterogeneous  execution model.}
As Figure~\ref{fig:representation_comparison} illustrates, another mismatch between monolithic data representation and SpBMM is that the entire Boolean matrix is forced through a homogeneous GPU execution model \cite{huang2020ge, gale2020sparse} that partitions and schedules matrices uniformly, making it impossible to apply fine-grained, sparsity-aware optimizations. Our key observation is that once matrix regions are encoded with distinct data structures, they no longer need to share the same execution path. Each data structure type induces its own memory-access pattern, so regions of the same type should be grouped together and mapped to matching processing paths, where type-specific optimizations can be applied to yield more efficient and regular memory access patterns. A good SpBMM design should not merely compress metadata more effectively; it should also enable a heterogeneous execution model in which regions are grouped and executed according to their local structure.

\subsection{System Overview}
\label{sec:insight}

The above opportunities
motivate us to rethink the optimization roadmap of SpBMM.
Rather than asking how to optimize execution for a single, monolithic format, we ask how to build an adaptive representation from complementary data structures, and then exploit the resulting optimization opportunities throughout the system.
Realizing this idea is challenging,
as it necessitates a shift from optimizing separate system components to the co-design of the data representation itself, the sparse Boolean workloads it must process,
the GPU execution and scheduling strategy it induces, and lower-level memory access patterns that make it efficient in practice.
The following sections detail how we realize this vision, beginning with an overview of  system architecture.

\begin{figure}[t!]
  \centering
  \vspace{-3mm}
  \includegraphics[width=0.85\linewidth]{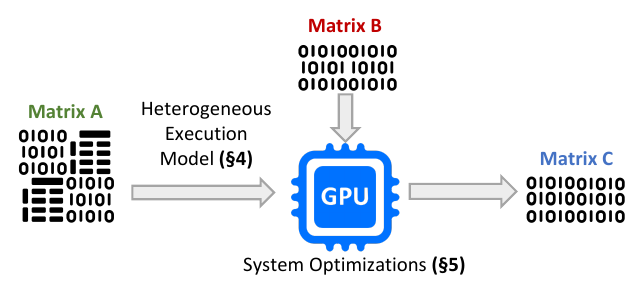}
  \vspace{-4mm}
  \caption{Overview of \sys SpBMM pipeline. }
  \Description{The overview of \sys}
  \vspace{-5mm}
  \label{fig:system_overview}
\end{figure}

Figure~\ref{fig:system_overview} illustrates the overall workflow of \sys. It begins by encoding matrix $A$ with an adaptive data representation (\S\ref{sec:structure}), which partitions the matrix into fine-grained element blocks and assigns each block the data structure best matched to its local sparsity. This encoding in turn drives the heterogeneous GPU execution model (\S\ref{sec:execution}), where element blocks of the same encoding type are grouped together, balanced across GPU execution units, and processed along type-specific execution paths. To streamline online execution, \sys uses bitmap as the default runtime interface for the dynamic input matrix $B$ and the output matrix $C$, avoiding repeated adaptive conversion during execution. On top of this execution model, \sys further introduces lower-level optimizations for memory layout and metadata indexing (\S\ref{sec:sysdesign}), improving data movement and access efficiency on the GPU. Together, these components form an efficient end-to-end SpBMM pipeline.

\section{\sys Data Representation}
\label{sec:structure}
In this section, we outline an adaptive data representation tailored
to Boolean matrices with highly skewed sparsity distribution.

\subsection{Adaptive Boolean Data Representation}

\begin{figure}[t]
  \centering
  \includegraphics[width=0.95\linewidth]{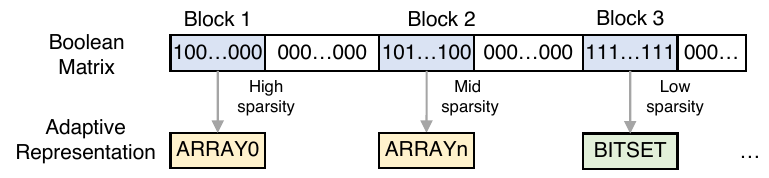}
  \vspace{-2mm}
  \caption{Constructing \sys data representation according to the sparsity level of fine-grained element blocks. }
  \vspace{-4mm}
  \Description{Element block construction diagram}
  \label{fig:representation_construction}
\end{figure}

To overcome the fundamental limitations of monolithic data structures, we propose an offset-bitmap adaptive data representation for sparse Boolean matrix processing. As shown in Figure \ref{fig:representation_construction},
the core principle is to abandon the one-size-fits-all approach, and instead (i) partition each matrix row into smaller, \rev{fixed-size} \textit{element blocks}, (ii) evaluate the local sparsity level of each element block, then (iii) encode each block using the most appropriate data structure. Specifically, the data structure is either selected from a family of offset-based formats (ARRAYn), with each variant optimized for a particular sparsity level, or adopts a bitmap-based format (BITSET), which is tailored for dense element blocks.

A core component of our adaptive representation is the element block. Rather than dividing a row into consecutive grid segments, element blocks are formed adaptively based on the actual positions of the non-zero elements.
Concretely, we partition each matrix row into fixed-length windows of size $l$ along the column dimension.
When traversing a row, the first non-zero element encountered at column index $c_{start}$ triggers the creation of an element block for the corresponding window.
The block then consolidates all non-zero elements whose column indices fall within that range.
Empty windows are skipped entirely, and the process proceeds to the next range that contains at least one non-zero element.
This on-demand generation mechanism is highly efficient, as it ensures that vast empty regions between clusters of non-zero elements generate no element blocks and therefore consume no space.

\begin{figure}[t]
  \centering
  \includegraphics[width=0.95\linewidth]{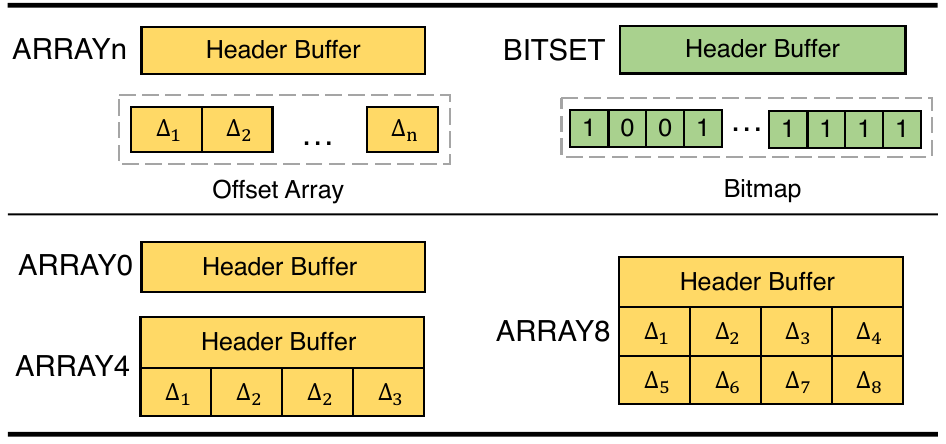}
  \vspace{-3mm}
  \caption{The format of offset-based data structure (ARRAYn)  and bitmap-based data structure (BITSET), with $\boldsymbol{l = 2^8}$.}
  \label{fig:array_bitset}
  \vspace{-4mm}
  \Description{ARRAYn and BITSET diagram. ARRAY0 and ARRAY4 as example}
\end{figure}

Based on the number of non-zero entries it contains, each element block is then compressed by one of two distinct families of data structures, namely ARRAYn and BITSET, detailed in Figure \ref{fig:array_bitset}.

\noindent\textbf{Offset-based representation (ARRAYn)}: For sparse blocks containing a small number of non-zero elements, we use a family of offset-based data structures. Instead of storing column indices of every non-zero elements in \texttt{uint32} format,
an element block only stores its first non-empty column index $c_{start}$ in the \textit{header buffer}. \rev{For each subsequent non-zero column index $c_i$, the \textit{offset array} stores $\Delta_i = c_i-\bar{c}_{start}$, where $\bar{c}_{start}$ denotes the beginning index of the element block.}
Storing offsets instead of \texttt{uint32} indices can significantly reduce space usage: Assume the range of an element block is $l=2^m$,
then we only need $m$ bits to represent any offsets within that element block. As a concrete example, if  $l=256=2^8$, then every offset can be stored in merely $m=8$ bits rather than $32$ bits. This, in turn, suggests that \textit{smaller element blocks (equivalently, smaller $m$) allows for a higher compression rate in offset arrays.}

To accommodate the local sparsity level of each block, we introduce a variable $n$ (as in ARRAY\textbf{n}) to represent the length of the block's corresponding offset array.
For example, if an element block contains 5 non-zero indices in total, an ARRAY4 structure will be used since it can store 1 start index (i.e., $c_{start}$) plus $n=4$ offsets.
For GPU efficiency, we pack the offset array to align with 32-bit boundary. To avoid wasting space or issuing misaligned loads, we require that the total number of offsets $n$ be an integer multiple of $32 / m$ (e.g., if $m=8$, then only ARRAY0, ARRAY4, ARRAY8, and so on can be used). This in turn implies that $m$ must be power-of-two divisor of 32 (e.g., 4, 8, 16).
Each element block is then padded to the nearest ARRAYn: if an element block has 6 non-zero indices, then it will be padded to ARRAY8.
Note that ARRAY0 is a special case where offset array has no entries at all, which means it can only hold one non-zero index---$c_{start}$ itself. Essentially, \textit{ARRAY0 is equivalent to the reduced CSR format used in CuBool.}

\noindent\textbf{Bitmap-based representation (BITSET)}: For denser blocks where the number of non-zeros exceeds a certain threshold $t$, we instead use a BITSET data structure. Similar with ARRAYn, every element block has a $c_{start}$ header buffer.
The difference lies in how the other non-zero indices are encoded: instead of relying on an offset array,
BITSET employs a bitmap with $l$ bits to represent all the $l$ columns in the block. \rev{A `1'} at the i-th bit position indicates a non-zero element at the corresponding i-th column. This dense format is exceptionally efficient for element blocks with many non-zero elements, as \rev{every element} can be stored in a single bit, marking a compression rate that surpasses offset-based counterparts.

\subsection{Configuring the Data Representation}

The effectiveness of the adaptive representation  depends heavily on the choice of parameters such as the range of element block $l=2^m$, and the switching threshold $t$ between ARRAYn and BITSET.
Rather than tuning them empirically, we derive suitable parameter values from a set of inherent design constraints. Concretely,
the choice of \sys parameter values should \textbf{(i)} \textit{maximize the overall space efficiency}, and \textbf{(ii)} \textit{avoid irregular memory access patterns}.

To start with, we know that the total space cost (in bits), for the ARRAYn (i.e., header buffer + offset array) and BITSET (i.e., header buffer + bitmap) data structures is given by:
$$
\text{cost(ARRAYn)}=
32 + m\cdot n,\ \ \ \
\text{cost(BITSET)} = 32 + 2^m
\label{eq:3_1_1}
$$
As such, to obtain the best space efficiency, ARRAYn is favored over BITSET if and only if $m\cdot n$ is smaller or equal to $2^m$, which translates to $n \leq 2^m/ m$.
Moreover, recall that to avoid irregular access patterns, $n$ must be a multiple of $32/m$, and $m$ must be a power-of-two divisor of 32.
We start by validating that $m=8$ is a viable choice, as the two constraints, (i) $n \leq 2^8 / 8 =32$ and (ii) $n$ must be an integer multiple of $32/8=4$ can hold at the same time.

We now proceed to rule out other choice of $m$.
For
$m<8$, we derive a contradiction.
Without loss of generality, let's assume $m=4$. In this case, we have $n \leq 2^4 / 4 =4$, while $n$ must also be an integer multiple of $32/4=8$. The only viable option is then $n=0$, which degenerates to the reduced-CSR case and violates our design goal of using offset arrays to handle sparse blocks.
For $m>8$,
constraint (ii) becomes the bottleneck: the bitmap length grows as $2^m$ bits, so $m = 16$ would yield per-block bitmaps of size $2^{16} = 65{,}536$ bits ($8 KB$), and ARRAYn blocks would likewise need thousands of bits of offsets. Such large per-block footprints force wide, non-coalesced memory accesses in the GPU, directly violating our requirement to avoid irregular access patterns.

\textit{In conclusion}, we adopt $m=8$ and element block length $l=256$ as they strike a good balance between space efficiency and access patterns
\rev{for the structure of modern GPU hardware.}
Under this choice, ARRAYn is competitive up to $n \le 32$, so we switch to BITSET when an element block contains more than $1 + 32$ non-zero indices, and set the switching threshold to $t = 33$.

\begin{figure}[t]
  \centering
  \vspace{-2mm}
  \includegraphics[width=\linewidth]{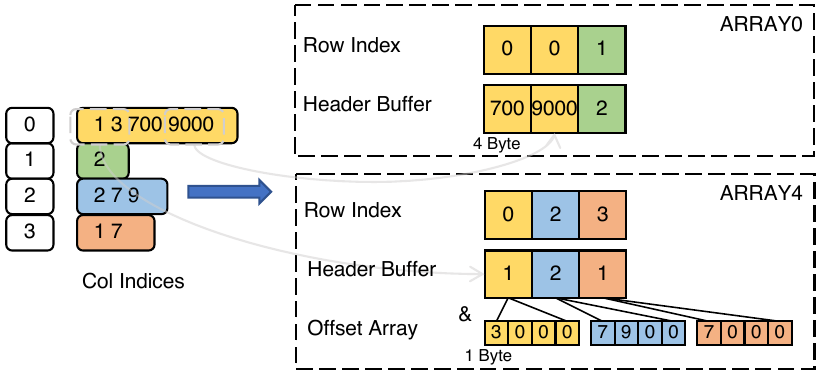}
  \vspace{-6mm}
  \caption{An example of constructing and grouping adaptive data representation, with element block length $\boldsymbol{l = 256}$.}
  \Description{Giving an example of Element block}
  \vspace{-4mm}
  \label{fig:block_grouping}
\end{figure}

\noindent\textbf{Example.} Figure \ref{fig:block_grouping} gives an example of the adaptive data representation with $l=256$ in action. The non-zero elements are partitioned into element blocks based on their column indices. Take Row 0 with non-zero elements at columns {1, 3, 700, 9000} as an example.  The elements at columns 1 and 3 fall within the same element block. Since there are two non-zero elements, the element block can be compressed into an ARRAY4 structure. The block's header buffer stores the full column index of column 1, and its offset buffer stores the offset of column 3. The elements at columns 700 and 9000 each start a new element block and, since they contain only a single non-zero element, they are both compressed into individual ARRAY0 structures.
Their respective headers store their full column indices (700 and 9000), and they have no associated offset array.

\section{\sys Execution Model}
\label{sec:execution}

In this section, we show how the proposed adaptive data representation enables an efficient GPU-based SpBMM execution model.

\subsection{Executing the Data Representation}
\label{sec:type_kernel}

Executing the adaptive representation efficiently on GPUs requires reconciling two seemingly conflicting goals.
On the one hand, the execution is inherently heterogeneous: different operands play different roles in SpBMM, and different element blocks may adopt different encodings according to local sparsity.
On the other hand, high GPU efficiency favors homogeneous processing paths that enable specialized optimization and avoid irregular access patterns.
To bridge this gap, \sys adopts a heterogeneous execution model with homogeneous processing paths: it preserves heterogeneity across operand roles and block types, while organizing execution so that each GPU kernel operates on a uniform class of work.

\noindent\textbf{Role-aware encoding.}
By default, the \sys adaptive representation is applied to matrix $A$, which in \rev{typical deployments is the larger and more frequently reused operand (e.g., a membership table)}, so that its encoding cost can be amortized across many SpBMM operations. The other operand $B$ (e.g., query requests) and the output $C$ are \rev{instead} represented in bitmap form by default, avoiding repeated adaptive conversion on the runtime path while preserving execution-friendly upstream and downstream interfaces for online processing.
This asymmetric encoding is a pragmatic design choice rather than a restriction of the framework. Crucially, it does not diminish the core benefit of the proposed adaptive representation: SpBMM already requires one operand to be traversed while the other is consulted during matching, so the main optimization leverage comes from adaptively encoding the traversed operand, while keeping the other operands lightweight to access and construct during execution.

\begin{figure}[t]
  \centering
  \vspace{-1mm}
  \includegraphics[width=\linewidth]{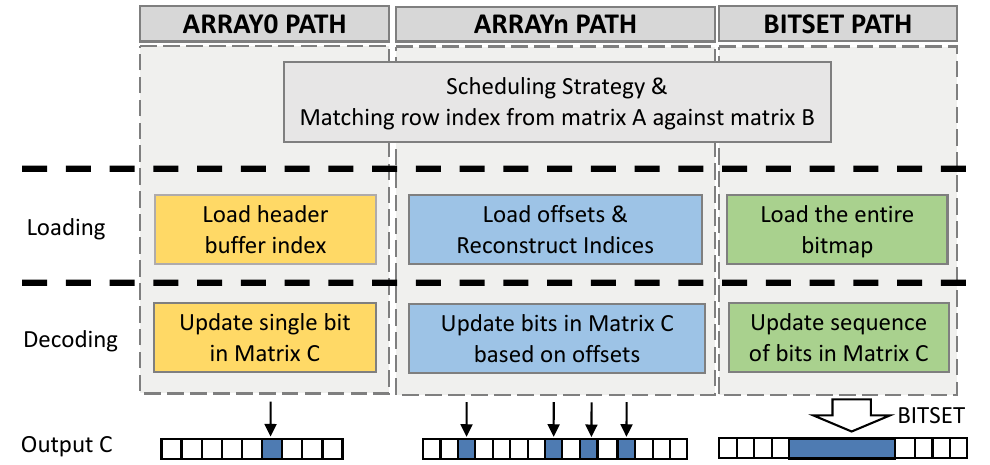}
  \vspace{-7mm}
  \caption{Type-based execution. All execution paths share the same matching process, but exploit distinct data structure layout for loading and updating optimizations.}
  \vspace{-5mm}
  \label{fig:type_kernel}
\end{figure}

\noindent\textbf{Type-based execution.}
Once the adaptive representation partitions the Boolean matrix $A$ into element blocks,
\sys groups blocks that use the same data structure into homogeneous \textit{element block pools}.
This organization naturally extends to GPU execution: different block types require fundamentally different load and update logic, ranging from single-bit materialization to offset reconstruction or bulk bitmap merge.
\sys therefore processes each pool with a dedicated kernel.
Doing so avoids mixing heterogeneous access patterns within the same execution path, while allowing the system to statically specialize memory-access widths, loop bounds, and decoding methods for each structure.
As illustrated in Figure~\ref{fig:type_kernel}, \sys maintains three classes of processing strategies (i.e., GPU kernels).
They share the same matching step, in which the row index of matrix $A$ is tested against the corresponding column of matrix $B$ to determine whether element blocks in that row contribute to the output.
They differ substantially, however, in how they load block contents and update matrix $C$.

\noindent\textbf{ARRAY0 processing.}
For element blocks that contain exactly one non-zero element, no offset array exists and no decoding is required.
On a match with matrix $B$, the processing logic
reads the 32-bit header buffer storing the column index of that single element,
and atomically updates the corresponding bit in the output bitmap $C$.
This is the lightest path but exhibits the most scattered writes.

\noindent\textbf{ARRAYn processing.}
For ARRAYn blocks containing multiple non-zero elements (i.e., $n>0$),
on a match with matrix $B$, the logic loads both the header buffer and the offset array
using a single vectorized memory transaction whose width is statically matched to the array size. For example, ARRAY4 stores
header buffer in 32 bits, then each of its 4 offsets in 8 bits,
resulting in a transaction width of 64 bits.
The logic then reconstructs full column indices by combining the aligned base header $\bar{c}_{start}$
(i.e., $c_{start}$ with the low 8 bits masked to zero)
with each 8-bit offset via a single bitwise OR operation.
This step is particularly lightweight because the aligned base and the offset occupy disjoint bit fields, avoiding a general addition-based index computation.
Each reconstructed index is then used to sequentially update the corresponding bit in the output matrix $C$.
Because the index reconstruction loop bounds are compile-time constants (i.e., fixed by the choice of $n$),
the compiler fully unrolls the loop and eliminates all branch overhead.

\noindent\textbf{BITSET processing.}
For dense element blocks that uses BITSET to encode non-zero elements,
the processing logic is fundamentally different.
On a match with matrix $B$,
the logic first loads both the header buffer and the entire 256-bit BITSET in a single vectorized memory transaction.
Crucially, because the BITSET block from matrix $A$ and the output matrix $C$ share the bitmap layout,
the loaded BITSET can then be written directly into $C$ at the granularity of 32-bit words, without any per-element decoding,
achieving the highest throughput per memory transaction among all strategies.

By processing each homogeneous pool with a dedicated GPU kernel,
\sys ensures that all threads within a launch follow the same processing path, maximizing instruction-level parallelism and eliminating warp divergence thereby improving execution efficiency. The type-based grouping also provides a natural unit for workload  scheduling and dispatching, as we show next.

\begin{figure}[t]
  \centering
  \includegraphics[width=\linewidth]{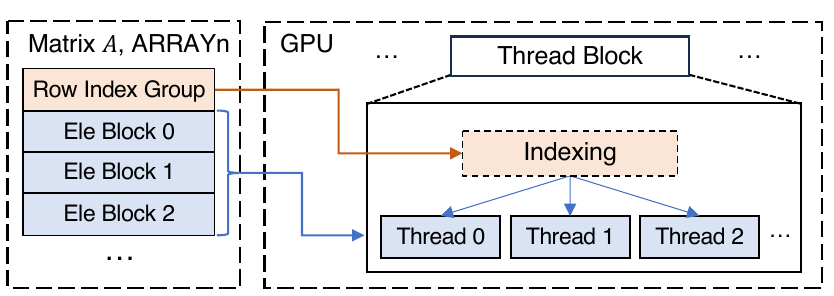}
  \vspace{-4mm}
  \caption{Traversal-based scheduling (\sys-T). Element blocks are evenly distributed across thread blocks.}
  \vspace{-4mm}
  \label{fig:type_execution}
\end{figure}

\subsection{Scheduling the GPU Execution}
\label{sec:type_execution}

Balanced workload distribution is critical for efficient GPU processing, as even type-specific execution can suffer from poor utilization when work is skewed across thread blocks.
By organizing element blocks into homogeneous pools and dispatching these pools sequentially, \sys exposes scheduling domains in which all scheduling units share the same data structure type and size.
This property enables effective workload balancing across GPU thread blocks.
Building on this foundation, \sys employs two complementary scheduling strategies tailored to different workload regimes.

\noindent\textbf{\sys-T: traversal-based scheduling.}
This strategy serves as the default scheduling mechanism of \sys.
It iterates over all element blocks of $A$, grouped by encoding type, and balances them across GPU thread blocks.
A naive row-based partitioning, as commonly adopted in general SpMM, is a poor fit here, because the number of element blocks varies substantially across rows, leading to severe workload imbalance in practice.
\sys-T therefore uses element blocks rather than matrix rows as the basic unit of workload distribution.
As illustrated in Figure~\ref{fig:type_execution}, \sys-T partitions the element blocks within each homogeneous pool into fixed-size chunks, and stores their corresponding row indices in compact \textit{row-index groups}.
Each chunk, together with its row-index group, is assigned to a single GPU thread block.

The execution then proceeds as follows.
First, each GPU thread block performs a constant-time metadata lookup to retrieve its assigned chunk of element blocks and row-index group.
Second, the row-index group is evenly partitioned and decoded across threads within the thread block, so that each thread is assigned the same number of element blocks.
For each element block, the processing logic uses its decoded row index to test the corresponding column in matrix $B$; on a match, it invokes the corresponding type-specific processing path.
Because every thread block and every thread processes the same number of element blocks, \rev{the workload is balanced} \textit{a priori}, without requiring complex runtime rescheduling.

\noindent\textbf{\sys-C: candidate-based scheduling.}
The default \sys-T path traverses all element blocks in $A$ regardless of whether their rows are activated by matrix $B$.
When $B$ is highly sparse, most of this work is wasted, because only a small subset of rows in $A$ can contribute to the final result.
To address this sparse-input regime, we introduce a candidate-based scheduling strategy that uses the non-zero entries of sparse matrix $B$ to directly filter the relevant rows of $A$.
By pruning irrelevant work before execution, \sys-C substantially reduces redundant memory accesses when the sparsity level of input is high.
The key challenge is that, although the non-zero elements of $B$ identify the relevant rows of $A$, the resulting workload remains highly irregular because different rows of $A$ may contain different numbers of element blocks.

\begin{figure}[t]
  \centering
  \includegraphics[width=0.97\linewidth]{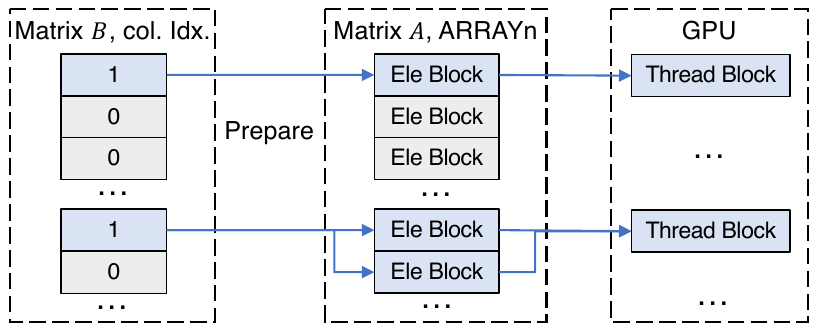}
  \vspace{-3mm}
  \caption{Candidate-based scheduling (\sys-C). It bypasses inactive rows in Matrix $A$ and only processes element blocks whose rows are activated by Matrix $B$.}
  \vspace{-5mm}
  \label{fig:row_index_bypassing}
\end{figure}

To address this problem, \sys-C adopts a two-phase design with a \textit{preparation phase} and an \textit{execution phase}, as shown in Figure~\ref{fig:row_index_bypassing}.
The preparation phase scans matrix $B$ to identify the active rows in $A$ and their corresponding element blocks, and partitions the resulting work into balanced task segments suitable for parallel GPU execution.
To accelerate this process, \sys-C maintains a precomputed prefix-sum array over the per-row element block counts of matrix $A$.
Consequently, the difference between adjacent prefix sums yields the total number of element blocks in a row, while the value at each index directly provides the memory offset needed to access them.
The execution phase then proceeds using the balanced task segments generated during preparation.
Each GPU thread block is assigned a single segment, from which it retrieves the corresponding row identifier, starting offset, and workload size, and then directly accesses the relevant element blocks in matrix $A$.
Threads within the block then collaboratively invoke the type-specific processing logic on those element blocks, without further matching against matrix $B$, because that step has already been completed during preparation.

\noindent\textbf{Choice of scheduling method.}
\label{sec:selector}
\sys-T and \sys-C are both effective scheduling strategies enabled by the \sys representation.
\sys-T leverages homogeneous element block pools to provide balanced traversal-based execution that performs robustly across workload regimes, while \sys-C can deliver greater speedup for very sparse inputs by using a preparation phase to directly prune the workload.
The two strategies are therefore inherently complementary: when matrix $B$ is sufficiently sparse, the pruning benefit of \sys-C outweighs its preparation cost; otherwise, \sys-T remains the preferable default path.
In practice, the better choice can be determined from workload characteristics, for example through lightweight offline profiling.
A fully automatic selector is orthogonal to our design and left to future work.

\section{\sys System Optimizations}
\label{sec:sysdesign}

In this section, we present  system optimizations built on top of the execution model that further accelerate data processing.

\subsection{Interleaved Memory Layout}
\label{sec:interleave}

During traversal-based scheduling, every GPU kernel repeatedly tests element blocks of $A$ against the input matrix $B$, so \sys caches the relevant slice of $B$ in each thread block's shared memory to amortize access latency. The effectiveness of this caching, however, directly depends on the underlying memory layout: Shared memory on modern GPUs is physically partitioned into many banks, and the threads in a warp achieve single-cycle read only when their accesses hit distinct banks; otherwise read accesses that collide on the same bank are serialized, introducing a bank conflict.
A row-major layout of matrix $B$ is particularly vulnerable to this pitfall: In \sys, the threads of a warp process the same column position of each row of $B$ in parallel.
Under row-major layout, the same word in consecutive rows is separated by a fixed stride equal to the row width, which for typical matrix dimensions aligns onto a small number of banks.
As a result, the threads in a warp collide on those banks, and their reads in the shared memory are serialized, negating the benefit of caching $B$ in the first place.

\begin{figure}[t]
  \centering
  \includegraphics[width=\linewidth]{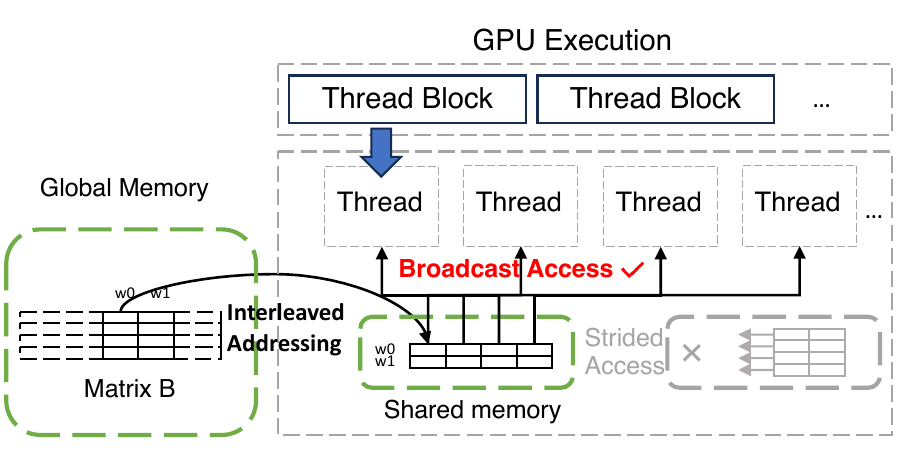}
  \vspace{-7mm}
  \caption{Interleaved addressing for matrix $B$.}
  \label{fig:interleaved_addressing}
  \vspace{-6mm}
\end{figure}

To overcome this, \sys organizes the cached slices of $B$ into a column-major layout during traversal-based scheduling. As illustrated in Figure~\ref{fig:interleaved_addressing},
instead of storing $B$ row by row, the words at the same column position across the rows are placed in consecutive shared-memory addresses when cached.
This inverts the access pattern of the row-major layout: the threads in a warp, which previously suffer from strided addresses and collide on the banks, now read those words from consecutive addresses that map to distinct banks.
Because traversal-based scheduling dispatches batches of element blocks whose row indices are clustered together, the column positions probed concurrently by a warp tend to be numerically close.
When all the threads of a warp happen to probe the same column word, the accesses collapse into a shared-memory broadcast, serving the entire warp with a single transaction; when they probe nearby but distinct column words, most accesses spread across distinct banks and still complete in a single transaction. Either way, the layout eliminates the conflict pathology described above.

\subsection{Aggregated Row Indexing }
\label{sec:index_compress}

In \sys's execution model, element blocks from different rows can be dispatched to the same thread, so every block must carry its own row index explicitly to remain addressable during execution.
Since this row index access happens on the critical path, retrieving it in full 32-bit format imposes non-trivial overhead on every kernel invocation, which is especially heavy for element blocks encoded by smaller data structures, e.g. ARRAY0. At the same time, the element blocks linked to each row index group often originate from nearby rows, leaving substantial redundancy to exploit. Building on this observation, we design an \textit{aggregated row index} that efficiently compresses row indices within each row index group, and enables on-the-fly decompression during kernel execution.

\begin{figure}[t]
  \centering
  \includegraphics[width=0.95\linewidth]{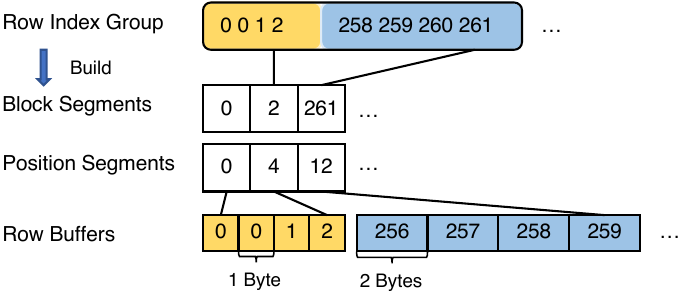}
  \vspace{-4mm}
  \caption{An example of aggregated row indexing.}
  \vspace{-4mm}
  \label{fig:row_aggregated_index}
\end{figure}

Within each element block pool, the sorted row indices are partitioned into
fixed-size row index groups.
Rather than storing every row index in the full 32-bit format, each row index group  only records a 32-bit \textit{base index} (the last row index in the previous group) in the block segment,
and stores the remaining indices as compact offsets relative to that base in a \textit{row buffer}.
The offset width in each row buffer is chosen according to the range it spans:
As illustrated in Figure~\ref{fig:row_aggregated_index}, the first row index group spans rows 0--2 (range $<2^8$), so its offsets occupy 1 Byte each; the second spans rows 2--260 (range $<2^{16}$), requiring 2 Bytes each.
A separate
\textit{position segments} array is then used to
track the beginning and ending address offsets of each row buffer.
For example, Figure~\ref{fig:row_aggregated_index}
shows that the second row buffer begins at Byte 4 and ends right before Byte 12.

At runtime, decompression leverages the aggregated row index structure to reconstruct row indices: each GPU thread (i) looks up the block and position segments to retrieve base index and locate the corresponding row buffer;
(ii) calculates row buffer offset width by diffing the current and next base indices;
(iii) fetches one offset from the row buffer, and (iv) reconstructs the full row index by adding the offset to the base index.
This light-weight decompression process ensures that aggregated row index reduces memory traffic without introducing substantial computation overhead.

\section{Evaluation}
\noindent\textbf{Implementation.}
We implement the \sys
sparse Boolean matrix processing pipeline in $\sim$25K lines of C++/CUDA code~\cite{adabool-repo}.
\sys is a production-grade system that has powered the attribute-based filtering service in our recommendation system for four years, serving millions of users daily.
In this section, we evaluate \sys on two GPU platforms: an NVIDIA RTX 4090 D GPU (Ada Lovelace, 24\,GB) and an NVIDIA A6000 GPU (Ampere, 48\,GB). Both servers run Ubuntu 22.04.5 with CUDA 11.8.

\noindent\textbf{Methodology.}
We compare \sys against representative general sparse matrix processing frameworks, including cuSPARSE~\cite{naumov2010cusparse}, Sputnik~\cite{gale2020sparse}, along with the specialized sparse Boolean matrix processing library cuBool~\cite{orachev2021spbla, wu2009batched}.
All open-sourced baselines are compiled from their latest available source releases for fairness.
To validate the generality of our system, we cover three common SpBMM patterns:
(i) sparse Boolean matrix-vector multiplication (SpBMV), where $B$ becomes a single vector $x$, representing single-query scenarios; (ii) small-scale
sparse Boolean matrix multiplication (SpBMM-S), where input matrix $B$ is substantially smaller than matrix $A$, representing batched-query scenarios; (iii) large-scale sparse Boolean matrix multiplication (SpBMM-L), where matrix $B$ is on the same scale as matrix $A$, representing complex matrix processing tasks.
The target workloads include a variety of synthetic micro-benchmarks, our production dataset \textit{AdsFilter}, and several public real-world datasets~\cite{leskovec2009community}.

\noindent\textbf{Evaluation questions.}
We organize the evaluation around four main questions: (1) How well does \sys perform compared with representative baselines?
(2) Why does \sys outperform existing systems?
(3) How robust and scalable is \sys across different SpBMM tasks?
(4) How much do \sys's individual design choices contribute to overall performance gains?
\vspace{-2mm}
\subsection{Micro-Benchmark}

\rev{In the micro-benchmark}, each group uses its own matrix scale to cover diverse real-world dimensions: SpBMV matches a vector $x$ against $A \in \mathbb{B}^{300K \times 300K}$;
\rev{SpBMM-S density sweeps match a narrow input matrix $B \in \mathbb{B}^{4 \times 300K}$ against $A \in \mathbb{B}^{300K \times 300K}$, simulating batched queries in our production workloads}; and SpBMM-L uses large-scale, square matrices $A$ and $B$ with row width $n$ up to 200K.
\rev{Unless otherwise stated, the default synthetic workloads use a Zipfian row-NNZ with skew $\alpha=1.0$ for SpBMV and SpBMM-S, and $\alpha=1.5$ for SpBMM-L.}
The density for any matrix or vector of dimension $m \times n$ is defined as $\textit{nnz}/(m \times n)$, where $\textit{nnz}$ is the total number of \rev{non-zero elements}.
In each group, we vary either the left operand's density or $A$'s density, probing both the effectiveness of our execution strategies and the robustness of the offset-bitmap adaptive representation. We use cuSPARSE, Sputnik, and cuBool as major baselines through all micro-benchmarking experiments, and we directly compare them against both our scheduling methods \sys-T (traversal-based scheduling) and \sys-C (candidate-based scheduling).

\begin{figure}[t]
  \centering
  \begin{subfigure}[b]{0.48\linewidth}
    \centering
    \includegraphics[width=\linewidth]{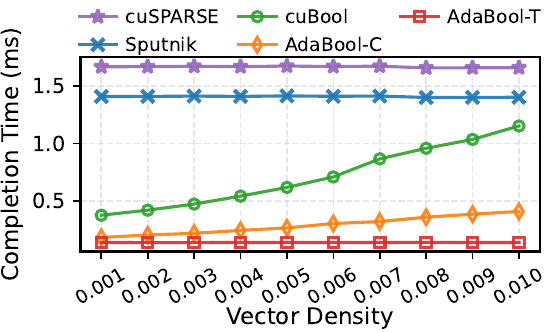}
    \caption{Varying $x$ density ($10^{-3}$--$10^{-2}$).}
    \label{fig:micro_spmv_a}
  \end{subfigure}
  \hfill
  \begin{subfigure}[b]{0.48\linewidth}
    \centering
    \includegraphics[width=\linewidth]{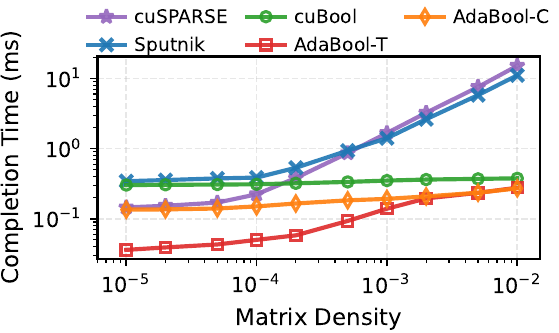}
    \caption{Varying $A$ density ($10^{-5}$--$10^{-2}$).}
    \label{fig:micro_spmv_b}
  \end{subfigure}
  \vspace{-4mm}
  \caption{Comparison of cuSPARSE, Sputnik, cuBool, and \sys on SpBMV: $A \in \mathbb{B}^{300K \times 300K}$.}
  \Description{SpBMV micro-benchmark under varying vector and matrix densities.}
  \label{fig:micro_spmv}
  \vspace{-4mm}
\end{figure}

\noindent\textbf{SpBMV.}
We compute the product $xA$ where $x$ is a sparse Boolean vector. Figure~\ref{fig:micro_spmv_a} varies the vector density from $10^{-3}$ to $10^{-2}$ while fixing $A$'s density at $10^{-3}$. The \sys-T method maintains a consistently low completion time of approximately 0.139~ms across all tested densities, outperforming cuSPARSE, Sputnik, and cuBool by up to 12.0$\times$, 10.1$\times$, and 8.3$\times$, respectively. This stability validates that  the traversal-based scheduling is largely insensitive to input sparsity level, as it always traverses all element blocks within $A$.
We observe that \sys-T outperforms \sys-C on SpBMV unless the vector density is ultra-sparse (e.g., at $10^{-4}$, \sys-C achieves 0.130~ms which is faster than \sys-T's 0.137~ms), because the preparation time of \sys-C cannot be amortized across rows in SpBMV given that $x$ is a single vector rather than a matrix.
\rev{
The gap narrows because cuBool's dominant cost, sorting query indices before each compute, shrinks with query density, but it does not invert: the residual cost is processing matrix $A$ itself, where \sys's compressed representation retains its advantage independent of query density.
}

\begin{figure}[t]
  \centering
  \begin{subfigure}[b]{0.48\linewidth}
    \centering
    \includegraphics[width=\linewidth]{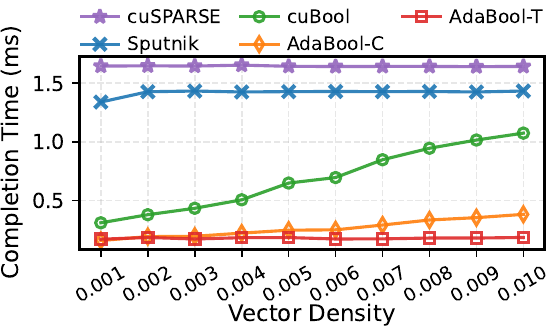}
    \caption{Uniform distribution.}
    \label{fig:spmv_distribution_uniform}
  \end{subfigure}
  \hfill
  \begin{subfigure}[b]{0.48\linewidth}
    \centering
    \includegraphics[width=\linewidth]{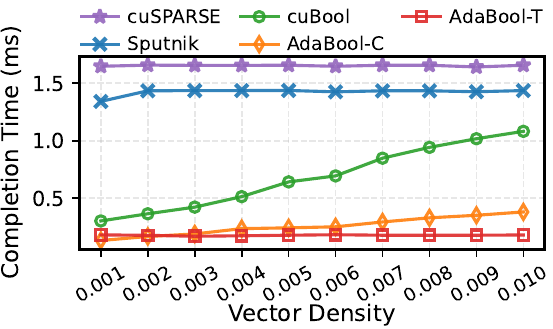}
    \caption{Gaussian distribution.}
    \label{fig:spmv_distribution_gaussian}
  \end{subfigure}
  \vspace{-4mm}
  \caption{Comparison of cuSPARSE, Sputnik, cuBool, and \sys on SpBMV: $A \in \mathbb{B}^{300K \times 300K}$ has target density $10^{-3}$}
  \Description{SpBMV completion time under uniform and Gaussian distributions.}
  \label{fig:spmv_distribution}
  \vspace{-2mm}
\end{figure}

Figure~\ref{fig:micro_spmv_b} fixes $x$'s density at $10^{-3}$ and sweeps $A$'s density from $10^{-5}$ to $10^{-2}$, directly evaluating the adaptive representation's robustness. As $A$ becomes denser, cuSPARSE's completion time increases by over $100\times$ (from 0.145~ms to 15.2~ms), while Sputnik increases by over $32.9\times$. In contrast, \sys-T increases by only $7.7\times$ (from 0.036~ms to 0.277~ms), achieving speedups of up to 54.7$\times$ and 40.0$\times$ over cuSPARSE and Sputnik, respectively. This resilience stems from the type-based execution on an adaptive data representation, which avoids the data movement overhead that afflicts monolithic CSR-based methods at higher densities.
\rev{Figures~\ref{fig:spmv_distribution_uniform} and~\ref{fig:spmv_distribution_gaussian} further show that \sys remains the fastest implementation across all vector $x$ densities under both uniform and Gaussian distributions, confirming that its advantage does not depend on Zipfian skew.}

\begin{figure}[t]
  \centering
  \begin{subfigure}[b]{0.48\linewidth}
    \centering
    \includegraphics[width=\linewidth]{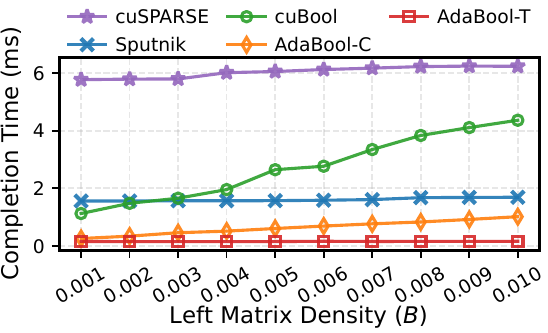}
    \caption{Varying $B$ density ($10^{-3}$--$10^{-2}$).}
    \label{fig:micro_bspmv_a}
  \end{subfigure}
  \hfill
  \begin{subfigure}[b]{0.48\linewidth}
    \centering
    \includegraphics[width=\linewidth]{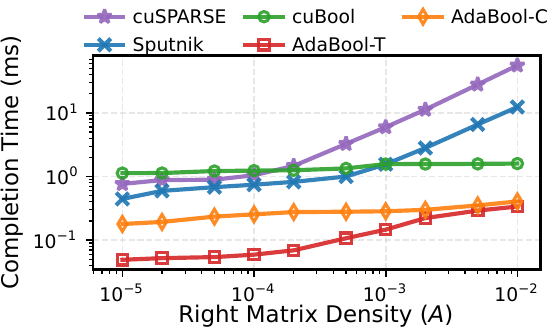}
    \caption{Varying $A$ density ($10^{-5}$--$10^{-2}$).}
    \label{fig:micro_bspmv_b}
  \end{subfigure}
  \vspace{-4mm}
  \caption{Comparison of cuSPARSE, Sputnik, cuBool, and \sys on SpBMM-S: $A \in \mathbb{B}^{300K \times 300K}$, $B \in \mathbb{B}^{4 \times 300K}$.}
    \Description{SpBMM-S micro-benchmark under varying left and right matrix densities.}
  \label{fig:micro_bspmv}
  \vspace{-3mm}
\end{figure}

\noindent\textbf{SpBMM-S.}
We evaluate small-scale matrix-matrix multiplication $BA$ with $B \in \mathbb{B}^{4 \times 300K}$ and $A \in \mathbb{B}^{300K \times 300K}$. Figure~\ref{fig:micro_bspmv_a} varies $B$'s density from $10^{-3}$ to $10^{-2}$ while fixing $A$'s density at $10^{-3}$. Across this sweep, \sys-T remains between 0.147 and 0.160~ms and outperforms cuSPARSE, Sputnik, and cuBool by up to 40.4$\times$, 10.8$\times$, and 27.4$\times$, respectively. Although \sys-C completion time increases from 0.256 to 1.014~ms as $B$ becomes denser, it remains faster than all three baselines. \sys-T is less sensitive because its traversal work is determined primarily by the fixed matrix $A$.

\begin{figure}[t]
  \centering
  \begin{subfigure}[b]{0.48\linewidth}
    \centering
    \includegraphics[width=\linewidth]{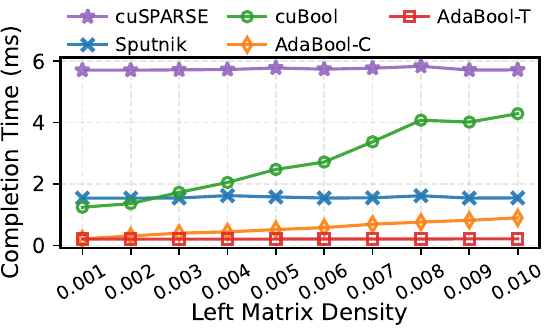}
    \caption{Uniform distribution.}
    \label{fig:spmm_s_distribution_uniform}
  \end{subfigure}
  \hfill
  \begin{subfigure}[b]{0.48\linewidth}
    \centering
    \includegraphics[width=\linewidth]{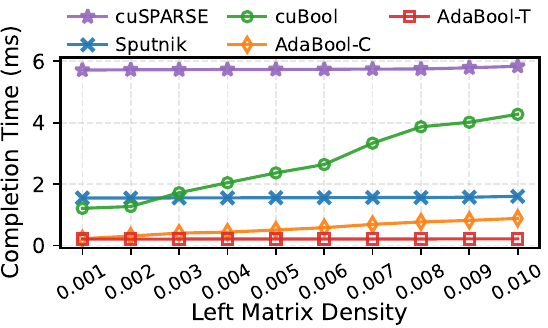}
    \caption{Gaussian distribution.}
    \label{fig:spmm_s_distribution_gaussian}
  \end{subfigure}
  \vspace{-4mm}
  \caption{Comparison of cuSPARSE, Sputnik, cuBool, and \sys on SpBMM-S: $A \in \mathbb{B}^{300K \times 300K}$, $B \in \mathbb{B}^{4 \times 300K}$.}
  \Description{SpBMM-S completion time under uniform and Gaussian  distributions.}
  \label{fig:spmm_s_distribution}
  \vspace{-2mm}
\end{figure}

Figure~\ref{fig:micro_bspmv_b} fixes $B$'s density at $10^{-3}$ and varies $A$'s density from $10^{-5}$ to $10^{-2}$. Across this sweep, cuSPARSE and Sputnik slow down by 73.1$\times$ and 27.9$\times$, respectively, whereas \sys-T increases by only 6.9$\times$, from 0.049 to 0.337~ms. At the highest density, \sys-T is 165.9$\times$, 36.8$\times$, and 4.7$\times$ faster than cuSPARSE, Sputnik, and cuBool, respectively. These results confirm that the adaptive representation remains effective for SpBMM-S as the right matrix becomes denser.
\rev{Figures~\ref{fig:spmm_s_distribution_uniform} and~\ref{fig:spmm_s_distribution_gaussian} further show that \sys-T consistently achieves the lowest end-to-end latency across all $B$ densities under both uniform and Gaussian distributions.}

\begin{figure}[t]
  \centering
  \begin{subfigure}[b]{0.48\linewidth}
    \centering
    \includegraphics[width=\linewidth]{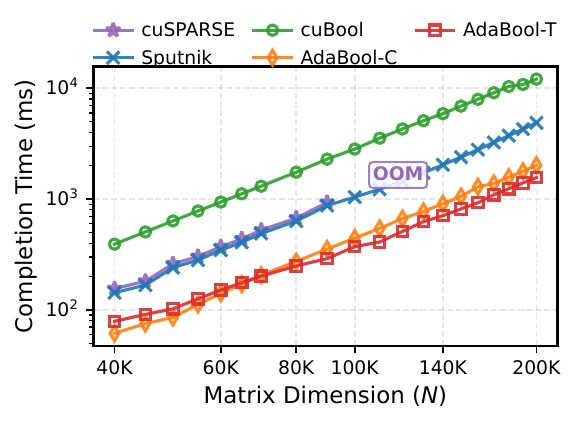}
    \caption{Varying Scale $n$ ($d = 10^{-3}$).}
    \label{fig:spmm_scale}
  \end{subfigure}
  \hfill
  \begin{subfigure}[b]{0.48\linewidth}
    \centering
    \includegraphics[width=\linewidth]{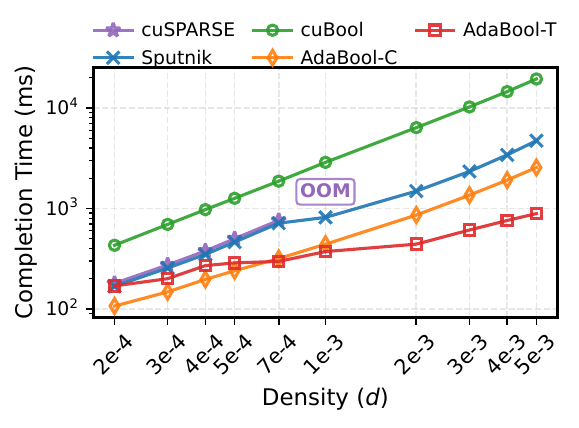}
    \caption{Varying density $d$ ($n = 100$K).}
    \label{fig:spmm_density}
  \end{subfigure}
  \vspace{-2mm}
    \caption{Comparison of cuSPARSE, Sputnik, cuBool, and \sys on SpBMM-L: $A, B \in \mathbb{B}^{n \times n}$.}
\Description{Large-scale square sparse Boolean matrix-matrix multiplication.}
  \label{fig:spmm_sensitivity}
  \vspace{-3mm}
\end{figure}

\noindent\textbf{SpBMM-L.}
We further evaluate large-scale matrix-matrix multiplication where $A$ and $B$ are equal-size square Boolean matrices~\cite{gale2020sparse}.
Figure~\ref{fig:spmm_scale} fixes the density at $10^{-3}$ and scales $n$ from 40K to 200K;
when $n$ is small and the preparation phase is cheap, \sys-C performs better than \sys-T;
as $n$ grows and the preparation becomes more expensive, \sys-T takes the lead beyond $n\approx 70$K, with the gap widening monotonically thereafter.
At $n = 90$K, \sys completes in 290~ms---$7.9\times$ and $3.2\times$ faster
than cuBool and cuSPARSE, respectively. cuSPARSE runs out of memory at
$n \geq 100$K, whereas \sys scales gracefully to $n = 200$K while retaining
a $7.7\times$ advantage over cuBool.

Figure~\ref{fig:spmm_density} fixes matrix size to $n = 100$K and varies density from $2\times10^{-4}$ to $5\times10^{-3}$.
\sys-C maintains the lowest completion time when density is less than $10^{-3}$, and \sys-T takes the lead when density is bigger than $10^{-3}$ where the preparation-based pruning becomes less effective.
cuSPARSE encounters out-of-memory failures beyond density $7\times10^{-4}$, while \sys scales gracefully. These results demonstrate that \sys's adaptive representation and execution strategies generalize effectively from the small to the large matrix multiplication setting.
They also empirically validate the complementary nature of the two execution strategies.
\rev{In terms of density distribution, with $n=100\mathrm{K}$ and target density $10^{-3}$, the uniform distribution yields average latencies of 390~ms for \sys-T and 515~ms for \sys-C, compared with 2219~ms for cuBool and 3983~ms for Sputnik; cuSPARSE runs out of memory.
Under the Gaussian distribution, \sys-T and \sys-C take 388~ms and 525~ms, respectively, compared with 2212~ms for cuBool and 4035~ms for Sputnik, while cuSPARSE again runs out of memory.}

\rev{\noindent\textbf{Cross-platform comparison.}
To further clarify the performance of \sys with frameworks designed for non-GPU hardware platforms, we run the
state-of-the-art CPU baseline SPALM~\cite{kim2026spalm} on an Intel Xeon Gold
5418Y with 16 threads and 128GB memory.
For SpBMV configuration with $A$'s density being $10^{-3}$, SPALM completes within 33.0~ms, 35.2~ms, 38.7~ms at $x$ densities $10^{-3}$, $5\times 10^{-3}$, and $10^{-2}$ respectively, while \sys-T completion time staying constantly below 0.14~ms.
For SpBMM-S configuration with $A$'s density being $10^{-3}$, SPALM completes within 496~ms, 501~ms, and 505~ms at $B$ densities $10^{-3}$, $5\times 10^{-3}$, and $10^{-2}$, respectively, while \sys-T completion time staying constantly below 0.16~ms.
For the SpBMM-L workload with density being $10^{-3}$ and scale being 100K, AdaBool completes in 371 ms, while SPALM takes 2634ms with more than 10$\times$ memory usage. When further scaled to the 200K setting, SPALM's lack of data representation optimization becomes fatal, directly resulting in an OOM error.

}

\subsection{Sources of Performance Advantage}
\label{sec:detail_analysis}

To answer what technical insights justify \sys's performance advantages over existing systems, we profile \sys against two representative baselines, cuSPARSE and cuBool.

\begin{figure}[t]
  \centering
  \begin{subfigure}[t]{0.493\linewidth}
    \centering
    \includegraphics[width=\linewidth]{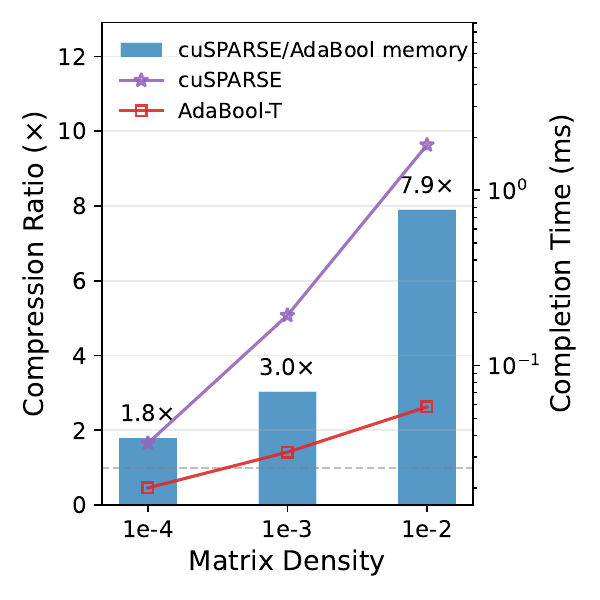}
      \vspace{-5mm}
    \caption{Memory comparison}
    \label{fig:detail_analysis_a}
  \end{subfigure}
  \hfill
  \begin{subfigure}[t]{0.493\linewidth}
    \centering
    \includegraphics[width=\linewidth]{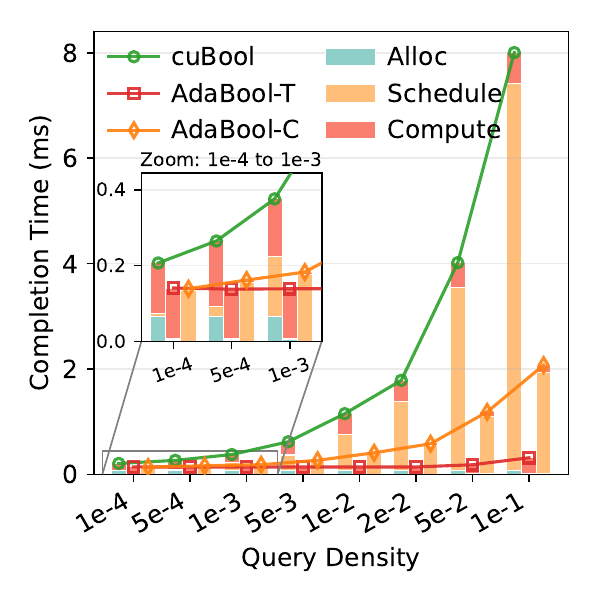}
      \vspace{-5mm}
    \caption{Phase breakdown}
    \label{fig:detail_analysis_b}
  \end{subfigure}
  \vspace{-2mm}
  \caption{Detail analysis of \sys's performance advantage over: (a)~cuSPARSE, (b)~cuBool.}

  \label{fig:detail_analysis_combined}
  \vspace{-3mm}
\end{figure}

\noindent\textbf{Speedup over cuSPARSE.}
Figure~\ref{fig:detail_analysis_a} plots the ratio of GPU memory consumed by cuSPARSE to that of \sys together with the latencies of cuSPARSE and \sys-T at matrix densities $10^{-4}$, $10^{-3}$, and $10^{-2}$ with fixed query density at $10^{-3}$.
The results clearly show that both \sys's memory consumption and completion time are lower than those of cuSPARSE.
Specifically, at $10^{-4}$ density, \sys achieves a $1.8\times$ memory saving ratio compared to cuSPARSE and a completion time of 0.020~ms compared to 0.036~ms for cuSPARSE.
At $10^{-2}$ density, \sys reaches a $7.9\times$ memory saving ratio over cuSPARSE, while its completion time remains only 0.058~ms, compared to 1.812~ms for cuSPARSE.
Notably, as matrix density increases, the performance gap between \sys and cuSPARSE widens in lockstep with the growing memory saving ratio, suggesting that memory reduction directly translates to completion time improvement.
This trend confirms that \sys's performance advantage over cuSPARSE stems largely from its adaptive data representation, which reduces memory accesses.

\noindent\textbf{Speedup over cuBool.}
As a Sparse Boolean matrix processing library, cuBool removes the numerical value access overhead; yet it remains significantly slower than \sys.
Figure~\ref{fig:detail_analysis_b} decomposes cuBool, \sys-T, and \sys-C completion time into three phases---\emph{Alloc} (\rev{allocating workspace and transfering data}), \emph{Schedule} (\rev{reordering, matching and preparation}), and \emph{Compute} (loading-and-updating)---across query ratios $10^{-4}$--$10^{-1}$ with fixed matrix density at $10^{-3}$.
For cuBool, the schedule phase dominates increasingly as the query becomes denser, growing from 4\% to 92\% of total completion time and driving a $39.5\times$ end-to-end increase, because cuBool must sort the query index during execution before each compute.
In contrast, \sys-T traverses the matrix $A$ directly during scheduling, making its preparation cost negligible and nearly constant across the entire range of query ratios.
Consequently, \sys-T is $1.5\times$ faster than cuBool at query ratio $10^{-4}$ but $25.5\times$ faster at $10^{-1}$, indicating that cuBool becomes increasingly dominated by query-dependent preparation overhead, whereas \sys largely removes this bottleneck through its heterogeneous execution model.

In summary, \sys simultaneously reduces the \emph{data volume} traversing the GPU memory hierarchy and eliminates the \emph{scheduling overhead} on the critical execution path.

\subsection{Production Workload: Attribute Filtering}
\label{sec:case1}

To evaluate the effectiveness of our design on real-world applications, we start by evaluating its effectiveness in our production environment.
We use AdsFilter, the open-sourced attribute filtering dataset from our recommendation system, where the filter matrix $A$ is extremely large and sparse (6.9M rows, 1.6M columns, with a density of only $6.4 \times 10^{-6}$), while each input query is characteristically dense,
with an average density around $\sim$55\%.
\rev{We include Themis, expressing filtering as relational operators on GPU, and Genie, a GPU inverted-index engine that answers queries by scanning and merging posting lists~\cite{hong2024themis,zhou2018generic}.}

Table \ref{tab:7_3_2} shows the single-query processing latency of \sys versus various baselines. Our \sys-T method demonstrates unequivocally superior performance, achieving an average latency of 0.158 ms, outperforming the best baseline, cuSPARSE, by a factor of 7.4$\times$ and the specialized cuBool library by 8.4$\times$. The high density of the query vectors makes this strategy ideal, as it fully leverages the traversal-based scheduling and adaptive representation of the large filter matrix, making performance largely independent of the query's structure. \sys-C on the one hand is dominated by the long and ineffective preparation phase due to ultra-dense queries and a large filter matrix, leading to higher latency.

We further compare against other attribute filtering methods:
Genie, a representative inverted list-based approach, records an average latency of 7.996 ms. Themis lags further behind at 11.391 ms, confirming the inefficiency of mapping SpBMM to relational operators.
These results directly validate the necessity of \sys adaptive representation and traversal-based scheduling. Furthermore, the stability of \sys-T method is particularly evident in its tail-latency performance, a critical metric for production systems with strict service-level objectives. At the 99th percentile, our method achieves a latency of only 0.235 ms, which is 5.3$\times$ faster than the next best approach, cuSPARSE (1.244 ms), ensuring highly predictable execution times even for worst-case queries \cite{gale2020sparse,han2018speeding}.

\begin{table}[t]
\centering
\caption{Query processing latency (ms) on AdsFilter.}
\vspace{-3mm}
\label{tab:7_3_2}
\begin{tabular}{lrrrrr}
\toprule
 & Avg & Std & P90 & P95 & P99 \\
\midrule
\textbf{Themis} & 11.391 & 10.080 & 34.900 & 35.700 & 36.500 \\
\textbf{Genie} & 7.996 & 6.228 & 7.800 & 7.826 & 13.044 \\
\textbf{cuBool} & 1.334 & 0.378 & 1.421 & 1.432 & 1.501 \\
\rev{\textbf{Sputnik}} & \rev{1.529} & \rev{0.073} & \rev{1.532} & \rev{1.533} & \rev{1.538} \\
\textbf{cuSPARSE} & 1.170 & \rev{\textbf{0.062}} & 1.204 & 1.209 & 1.244 \\
\textbf{\sys-T} & \textbf{0.158} & \rev{0.082} & \textbf{0.203} & \textbf{0.207} & \textbf{0.235} \\
\textbf{\sys-C} & 6.498 & 1.494 & 8.597 & 8.773 & 9.524 \\
\bottomrule
\end{tabular}
\vspace{-2mm}
\end{table}

\begin{figure}[t]
  \includegraphics[width=0.95\linewidth]{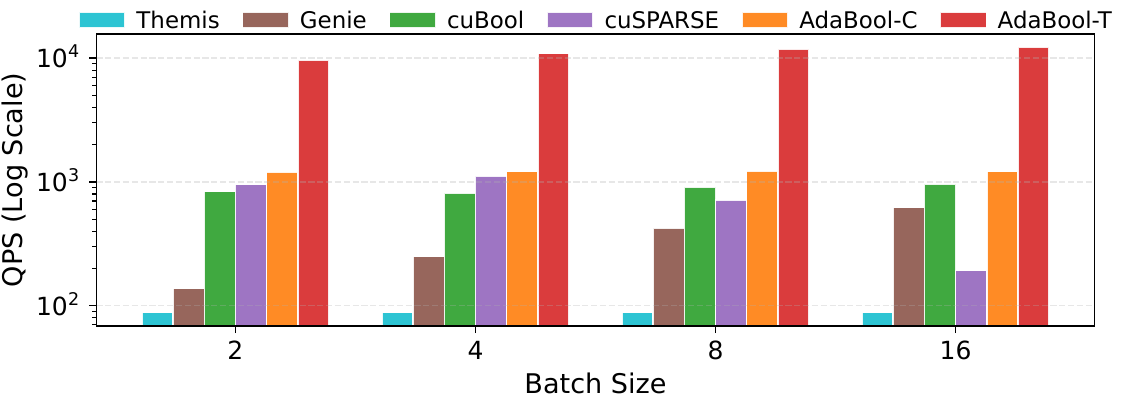}
  \vspace{-4mm}
  \caption{The throughput of processing attribute-based filtering under various batch sizes on AdsFilter.}
  \Description{This figure compares the throughput (QPS) of four different methods (cuSPARSE, cuBool, \sys-C, and \sys-T) across various batch sizes (2, 4, 8, 16). The bar chart uses a logarithmic y-axis to illustrate the performance differences, showing that \sys-T method achieves significantly higher throughput compared to other approaches.}
  \vspace{-5mm}
  \label{fig:industry_throughput}
\end{figure}

To further assess system scalability, we evaluate the throughput of our design with batched query processing. Figure \ref{fig:industry_throughput} exhibits the throughput comparison measured in queries per second (QPS) across all methods. Our \sys-T method exhibits exceptional scalability, achieving a peak throughput of 12,279 QPS at batch size 16, which represents a 63.9$\times$ improvement over cuSPARSE and a 12.8$\times$ speedup over cuBool at the same batch size. Themis plateaus at 87 QPS, illustrating the relational model's inability to handle batched matrix workloads. Genie shows moderate scaling, peaking at 619 QPS, but still falls orders of magnitude short of \sys-T.

Notably, as batch size increases, our \sys-T method's throughput continues to improve, demonstrating the effectiveness of our indexing and execution strategy in maintaining balanced workloads across thread blocks. In contrast, cuSPARSE exhibits severe performance degradation at larger batch sizes, with throughput collapsing from 1,113 QPS to merely 192 QPS. This instability mirrors the observations reported in prior work on sparse matrix operations \cite{gale2020sparse}, where monolithic data structures fail to maintain consistent performance under diverse workload conditions. The \sys-C method maintains modest but stable throughput around 1,200 QPS across all batch sizes, confirming it is bottlenecked by the dense query vectors characteristic of this application. These results collectively validate that our integrated design, combining the offset-bitmap adaptive representation with type-based execution model, consistently delivers both superior performance and robust scalability for real-world, large-scale filtering workloads.

\noindent\textbf{\sys construction overhead.}
On AdsFilter, \sys constructs its data representation in 221\,ms, 2.0$\times$ faster than cuSPARSE (444\,ms) and nearly two orders of magnitude faster than cuBool (16{,}688\,ms).
\rev{
The low construction overhead is achieved through a highly parallel, two-pass generation mechanism. The first pass simply counts elements to determine the local sparsity and representation type of each element block, introducing negligible overhead relative to other baselines.
The second pass then writes these type-specific structures directly into exactly sized, pre-allocated buffers with massive parallelism.
Furthermore, this internal efficiency is complemented by AdaBool's lightweight interface, which directly ingests contiguous input pairs to minimize initial data-loading costs.}
In many real-world scenarios, matrix $A$ encodes stable entities and is therefore rarely updated, so construction is a one-time cost amortized over many queries, introducing negligible overhead.
Regardless of this assumption, the construction overhead of \sys is lower than that of representative baselines.

\noindent\textbf{Cross-\rev{GPU} Portability.}
To demonstrate the generalizability of \sys's performance, we replicate the AdsFilter experiment on the NVIDIA A6000 platform, which is one generation older than the RTX 4090\,D.
Table~\ref{tab:cross_platform} reports the single-query average latency on both platforms, where
\sys shows consistent performance gains.
Interestingly,
\sys achieves a 16.3$\times$ speedup over cuSPARSE on the A6000 GPU, which is two times higher than the speedup on the RTX4090D. This is because cuSPARSE relies on hardware architecture-specific optimizations enabled by the newer generation GPU, while \sys is driven by fundamental representation-level optimization rather than hardware-specific features.
These results confirm that our approach can generalize seamlessly across GPU generations.

\begin{table}[t]
\centering
\caption{Cross-\rev{GPU} comparison on production AdsFilter dataset (single-query average latency in ms).}
\vspace{-3mm}
\label{tab:cross_platform}
\begin{tabular}{lrrrr}
\toprule
\textbf{GPU} & \textbf{cuBool} & \textbf{cuSPARSE} & \textbf{\sys} & \textbf{Speedup} \\
\midrule
RTX 4090\,D & 1.334 & 1.170 & \textbf{0.158} & 8.4$\times$/7.4$\times$ \\
RTX A6000       & 1.839 & 3.501 & \textbf{0.215} & 8.6$\times$/16.3$\times$ \\
\bottomrule
\end{tabular}
\vspace{-2mm}
\end{table}

\subsection{Public Benchmarks: Real-World Datasets}

We further evaluate \sys\ on three real-world public datasets---Email-Enron~\cite{leskovec2009community}, pdb1HYS~\cite{5348797}, and Human-Gene1~\cite{Belcastro2011TranscriptionalGN}---spanning nearly three orders of magnitude in density and size.
\rev{Specifically, Email-Enron is a 36{,}692 square matrix with 367{,}662 nonzeros and an average density about $10^{-4}$;
pdb1HYS is a 36{,}417 square matrix with an average density about $10^{-3}$;
and Human-Gene1 is a 22{,}283 square matrix with an average density about $10^{-2}$. The coefficients of variation of per-row density are 3.60, 0.27, and 1.27, respectively.}

\begin{figure}[t]
  \includegraphics[width=0.88\linewidth]{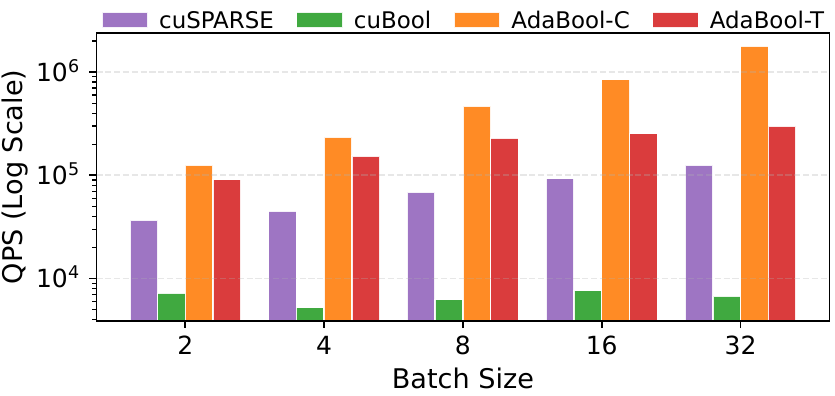}
  \vspace{-4mm}
  \caption{Throughput (QPS) on \textit{Email-Enron} query processing under various batch sizes.}
  \Description{Bar chart comparing throughput of cuBool, cuSPARSE, \sys-T, and \sys-C on Email-Enron across batch sizes 1--32.}
  \label{fig:case2_throughput}
  \vspace{-3mm}
\end{figure}

\vspace{1mm}
\noindent\textbf{Query Processing.}
We evaluate \sys on the query processing workload using the \textit{Email-Enron} dataset with sparse query vectors (average density $\approx 10^{-3}$).
As shown in Table~\ref{tab:case2_single}, \sys-C achieves an average latency of 0.014\,ms, outperforming cuBool by 12.3$\times$ and cuSPARSE by 2.7$\times$.
In contrast to the AdsFilter dataset where dense queries favored \sys-T, the sparse query vectors here make \sys-C the better choice, validating its unique advantage.
Figure~\ref{fig:case2_throughput} shows the batch throughput on \textit{Email-Enron}: \sys-C reaches over 1M QPS at batch size 32, a \textbf{264$\times$} speedup over cuBool.

\begin{table}[t]
\centering
\vspace{-1mm}
\caption{Query processing latency (ms) on \textit{Email-Enron}.}
\vspace{-3mm}
\label{tab:case2_single}
\begin{tabular}{lrrrrr}
\toprule
 & Avg & Std & P90 & P95 & P99 \\
\midrule
\textbf{cuBool}    & 0.172 & 0.050 & 0.179 & 0.182 & 0.188 \\
\textbf{cuSPARSE}  & 0.038 & 0.152 & 0.039 & 0.040 & 0.044 \\
\textbf{\sys-T}    & 0.025 & 0.092 & 0.023 & 0.023 & 0.025 \\
\textbf{\sys-C}    & \textbf{0.014} & \textbf{0.004} & \textbf{0.018} & \textbf{0.019} & \textbf{0.023} \\
\bottomrule
\end{tabular}
\vspace{-2mm}
\end{table}

\vspace{1mm}
\noindent\textbf{Matrix Multiplication.}
Table~\ref{tab:case2_spbmm_realworld} summarizes the matrix multiplication performance of various baselines on the three real-world datasets. Both \sys scheduling methods outperform all major baselines, and \sys-C achieves the lowest completion time on every dataset, with 0.017--0.021\,ms latency overall. Relative to the fastest representative baseline on each dataset, \sys-C delivers 4.9$\times$--32.3$\times$ speedups, showing that the adaptive data representation and heterogeneous execution model remain effective across different realistic matrix sizes and sparsity levels.

\begin{table}[t]
\centering
\caption{Completion time (ms) on real-world datasets.}
\vspace{-2mm}
\label{tab:case2_spbmm_realworld}
\begin{tabular}{lrrr}
\toprule
 & pdb1HYS & Human-Gene1 & Email-Enron \\
\midrule
\textbf{cuBool}   & 0.742 & 0.678 & 0.766 \\
\textbf{cuSPARSE} & 0.209 & 1.425 & 0.089 \\
\textbf{Sputnik}  & 0.199 & 1.340 & 0.084 \\
\textbf{\sys-T}   & 0.028 & 0.033 & 0.026 \\
\textbf{\sys-C}   & \textbf{0.018} & \textbf{0.021} & \textbf{0.017} \\
\bottomrule
\end{tabular}
\vspace{-2mm}
\end{table}

\subsection{Ablation Study}
\begin{table}[t]
\centering

\caption{GPU Memory consumption on real-world datasets.}
\vspace{-3mm}
\label{tab:memory_realworld}
\resizebox{\linewidth}{!}{%
\begin{tabular}{lrrrr}
\toprule
\textbf{Method} & \textbf{AdsFilter} & \textbf{Email-Enron} & \textbf{pdb1HYS} & \textbf{Human-Gene1} \\
\midrule
COO & 783.3 MB & 4.2 MB & 49.7 MB & 282.3 MB \\
CSR & 548.5 MB & 2.9 MB & 33.3 MB & 188.3 MB \\
cuBool & 293.4 MB & 1.7 MB & 17.0 MB & 94.4 MB \\
cuSPARSE  & 703.6 MB & 4.2 MB & 36.1 MB & 198.4 MB \\
\textbf{\sys} & \textbf{149.1 MB} & \textbf{1.1 MB} & \textbf{3.5 MB} & \textbf{25.2 MB} \\
\bottomrule
\end{tabular}%
}
\end{table}

To further validate the effectiveness of \sys's core components, we conduct a series of ablation studies on its adaptive data representation, execution model, and system optimizations.

First, we evaluate the memory efficiency of our offset-bitmap adaptive representation.
Table~\ref{tab:memory_realworld} compares the GPU memory footprint of \sys against COO, CSR, cuSPARSE and cuBool on four real-world datasets.
As a specialized Boolean library, cuBool inherently avoids the unnecessary value array overhead of standard CSR formats, effectively serving as a highly memory-efficient reduced CSR baseline. Yet, even against this strict baseline, \sys further reduces memory consumption to just 0.51$\times$ on the AdsFilter dataset and as low as 0.21$\times$ on pdb1HYS. This demonstrates the superior effectiveness of our offset-bitmap adaptive encoding. The memory savings become even more pronounced with increased matrix density: at higher densities, more blocks qualify for BITSET encoding (requiring only 1~bit per index compared to 32~bits in standard CSR), yielding dramatic compression.

Notably, the compression ratio varies across datasets due to their distinct sparsity profiles. For pdb1HYS and Human-Gene1, which exhibit moderate-to-high local density, a large fraction of element blocks qualify for BITSET or ARRAY32 encoding, yielding compression ratios of 0.21$\times$ and 0.27$\times$ relative to cuBool, respectively. In contrast, AdsFilter and Email-Enron are dominated by short posting lists (ARRAY0/ARRAY4), where per-element header overhead is higher, yet \sys still achieves 0.51$\times$ and 0.65$\times$ of cuBool's footprint. This diversity confirms that monolithic data structures cannot effectively accommodate the highly skewed sparsity distributions inherent in real-world Boolean matrices, and justifies why the \sys must comprise ARRAYn and BITSET.

\begin{table}[t]
  \rev{%
  \small
  \setlength{\tabcolsep}{3pt}
  \caption{Per-format composition of \sys on AdsFilter.}
  \vspace{-3mm}
  \label{tab:fmt}
  \begin{tabular}{@{}lrrrrrr@{}}
    \toprule
    Metric & ARRAY0 & ARRAY4 & ARRAY8 & ARRAY16 & ARRAY32 & BITSET \\
    \midrule
    Block & 77.8\% & 16.1\% & 2.4\% & 1.5\% & 1.0\% & 1.2\% \\
    Element & 24.9\% & 14.0\% & 5.5\% & 6.3\% & 7.4\% & 41.9\% \\
    Memory & 59.0\% & 20.6\% & 4.4\% & 4.6\% & 5.0\% & 6.4\% \\
    \bottomrule
  \end{tabular}
  }
  \vspace{-3mm}
\end{table}

\begin{table}[t]
  \rev{%
  \small
  \setlength{\tabcolsep}{1.5pt}
  \caption{Block composition (\%) on other real-world datasets.}
  \vspace{-3mm}
  \label{tab:fmt_memory_realworld}
  \begin{tabular}{@{}lrrrrrr@{}}
    \toprule
    Dataset & ARRAY0 & ARRAY4 & ARRAY8 & ARRAY16 & ARRAY32 & BITSET \\
    \midrule
    Email-Enron & 63.1 & 30.1 & 3.9 & 1.9 & 0.7 & 0.3 \\
    pdb1HYS & 0.7 & 13.2 & 14.8 & 12.2 & 20.5 & 38.5 \\
    Human-Gene1 & 15.4 & 31.1 & 13.8 & 13.0 & 11.1 & 15.7 \\
    \bottomrule
  \end{tabular}
  }
  \vspace{-3mm}
\end{table}

\rev{Table~\ref{tab:fmt} characterizes AdsFilter from three complementary perspectives: ``Block'' is the fraction of the final encoded blocks assigned to each format, ``Element'' is the fraction of the original matrix's column indices represented by each format, and ``Memory'' is the fraction of GPU memory consumed by each format.
Table~\ref{tab:fmt_memory_realworld} extends the block breakdown to the public datasets.
These distinct distributions show that real workloads exercise the full format family and that no single format suffices.

}

\begin{figure}[t]
  \includegraphics[width=0.9\linewidth]{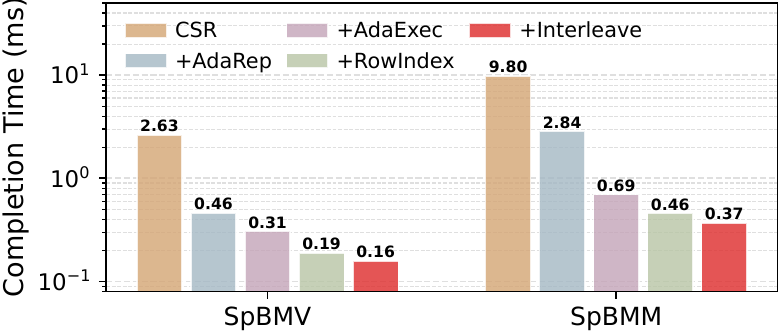}
  \vspace{-4mm}
  \caption{Ablation study of \sys system components.}
  \Description{Ablation study of \sys system components across five progressive configurations.}
  \label{fig:ablation_breakdown}
  \vspace{-5mm}
\end{figure}

Figure \ref{fig:ablation_breakdown} provides a more detailed breakdown of how each design component incrementally contributes to performance on the attribute-based filtering dataset. \textit{CSR} utilizes the CSR format with a naive execution strategy. Subsequent configurations progressively integrate our techniques: employing the offset-bitmap adaptive representation (\textit{+AdaRep}), adding the \sys execution \rev{model} (\textit{+AdaExec}), enabling aggregated row indexing (\textit{+RowIndex}), and finally enabling the complete system with interleaved addressing for memory layout (\textit{+Interleave}). The results show that the performance gains of \sys arise from the holistic co-design of data representation, execution strategy, and memory layout rather than from any single optimization in isolation.
Comparing \textit{CSR} against \textit{+Interleave} reveals dramatic overall speedups of 16.7$\times$ for SpBMV (2.63\,ms $\to$ 0.16\,ms) and 26.6$\times$ for SpBMM (9.80\,ms $\to$ 0.37\,ms), showing that the same architecture benefits both individual lookups and batched query processing.

The most substantial gain comes from adopting the offset-bitmap adaptive representation (\textit{+AdaRep}), which delivers a 5.7$\times$ SpBMV speedup over \textit{CSR} and a 3.5$\times$ speedup for SpBMM. By adaptively encoding each row block as ARRAYn or BITSET according to local density, \textit{+AdaRep} dramatically reduces the per-entry memory traffic each kernel must process compared to monolithic CSR format. The \sys execution model in \textit{+AdaExec} then contributes a further 1.5$\times$ improvement for SpBMV and a striking 4.1$\times$ improvement for SpBMM by tailoring kernel dispatch and memory access patterns---vectorized loads, shared-memory bitmap caching---to each encoding type, exploiting the highly skewed sparsity distribution in this workload. Adding aggregated row indexing in \textit{+RowIndex} provides a further 1.6$\times$ additional improvement, as the compressed row indices reduce global memory traffic during kernel execution.
Finally, interleaved addressing for memory layout (\textit{+Interleave}) is particularly impactful in the SpBMM setting, where multiple queries amplify memory traffic: it reduces SpBMM completion time from 0.46\,ms to 0.37\,ms by transforming irregular accesses into coalesced memory transactions that better utilize GPU bandwidth.

\vspace{-1mm}

\section{Related Work}

\textbf{Sparse matrix multiplication.} The optimization of sparse matrix multiplication on GPUs is a well-established research area, with NVIDIA's cuSPARSE and the open-source Sputnik library emerging as two of the most representative and state-of-the-art solutions \cite{naumov2010cusparse, gale2020sparse}. As the official, closed-source library offered by GPU vendors, cuSPARSE provides a highly optimized, industrial-grade baseline that is broadly used in production environments \cite{naumov2010cusparse}. In parallel, Sputnik, a project driven by the research community, has likewise emerged as a practical benchmark thanks to its outstanding performance. Concretely,
Sputnik has seen direct integration into popular frameworks like PyTorch \cite{gale2020sparse, paszke2019pytorch}.

Besides cuSPARSE and Sputnik, many other strategies have been explored to optimize SpMM. Approaches such as spECK and GE-SpMM focus on improving memory efficiency of the CSR data representation through workload repartitioning techniques \cite{parger2020speck, huang2020ge, anzt2020load}.
A different line of work proposes alternative data representation formats; for instance, the ELLPACK-Block format adaptively partitions the matrix into uniform blocks to improve load balance for SpMV, though its efficacy is limited to matrices with moderate and relatively uniform row densities \cite{chen2023gpu, isupov2021implementation, flynn2025spmm, shanbhag2022tile}.

Recent research has further investigated the use of specialized hardware units like Tensor Cores for sparse computation in GNNs \cite{wang2023tc, xia2025voltrix, xiang2025cutespmm, fan2024dtc, 11209980, hu2022tcudb}. While these methods offer compelling optimizations \rev{and leverage adaptive representation}, their broader applicability is constrained by the underlying hardware accelerators. Take Tensor Cores as an example: On the one hand, they are built to speed up computations on small, dense matrix tiles, which makes them poorly suited to SpBMM, where at least one input matrix is quite large and sparse. On the other hand, they are specialized for accelerating algorithmic computations, which mismatches with SpBMM as the latter purely relies on logical operations and by design avoids algorithmic computations.

\noindent \textbf{Domain-specific efforts.} An important body of research focuses on accelerating sparse computations for specific application domains \cite{wang2024cubit, li2024tengraph, hong2024themis, hu2024fast}. The most closely related work to our own is \textit{cuBool}, which is a state-of-the-art SpBMM processing library on GPUs and our primary baseline for comparison \cite{orachev2021spbla}. In the domain of graph analytics, Bit-GraphBLAS and GraphBLAS propose kernels to accelerate matrix-centric graph processing \cite{davis2023algorithm, chen2022bit, yang2022graphblast}.
A parallel line of research in information retrieval has also leveraged GPUs for Boolean-like operations; for instance, Griffin and LiNR present industrial-scale systems for query processing in retrieval systems \cite{borisyuk2024linr, liu2018griffin, zhang2025faaqp, ding2009using, mansour2024revisiting}.
Similarly, for database and index domain, Themis and Genie optimize query execution through GPU acceleration respectively \cite{hong2024themis, hu2024fast, zhou2018generic, zhu2024hyperblocker}, and recent work has further explored GPU-accelerated analytics on compressed data \cite{shanbhag2020study, huang2025gpu}. However, primary contributions of these latter systems lie in application-level architecture, scheduling, or theoretical bounds, rather than fundamental optimization to the SpBMM backend itself~\cite{10.1145/3571157}.

\rev{Another line of research focuses on optimizing sparse matrix multiplication on CPUs. Libraries such as Intel MKL and SuiteSparse have been developed to leverage the CPU's computational capabilities \cite{intel2023mkl, davis2004suitesparse}. Other works also implement adaptive data structures and algorithms to improve performance on CPUs \cite{kernert2016topology, kim2026spalm}. However, these CPU-focused optimizations are not directly applicable to GPU architectures due to fundamental differences in memory hierarchy, parallelism, and execution models.
}

\vspace{1.5mm}

\section{Conclusion}
Sparse Boolean Matrix Multiplication (SpBMM) is essential for many data processing tasks.
This paper makes the observation that achieving high-performance SpBMM on GPUs requires moving beyond monolithic data representations toward an adaptive paradigm that matches the highly skewed sparsity distribution of real-world Boolean matrices.
We introduce \sys, a SpBMM system that leverages an adaptive data representation
to partition matrices and encode each region differently based on local sparsity levels.
\sys further incorporates a heterogeneous execution model
equipped with role-aware encoding, type-based execution, and block-level scheduling mechanisms,
along side a set of system optimizations to fully unleash the performance benefits on modern GPUs.
Extensive evaluations demonstrate that \sys achieves significant performance improvement over representative baselines across diverse workloads.

\newpage
\bibliographystyle{ACM-Reference-Format}
\bibliography{sample}

\end{document}